\documentclass[aps, pra, twocolumn, superscriptaddress, amsmath, amssymb, noeprint]{revtex4-2}

\usepackage{graphicx}
\usepackage{dcolumn}
\usepackage{bm}
\usepackage{here}

\usepackage{physics}
\usepackage{siunitx}
\usepackage{comment}
\usepackage{makecell}
\usepackage{multirow}
\usepackage[svgnames]{xcolor}
\usepackage[colorlinks,citecolor=Blue,linkcolor=Blue,urlcolor=Blue,linktocpage,unicode]{hyperref}

\begin{document}

\title{Analyzing the sensitivity of an atom interferometer \\ with a phase-modulation readout scheme}

\author{Takuya~Kawasaki}
  \email{takuya.kawasaki@phys.s.u-tokyo.ac.jp}
  \altaffiliation[Present address: ]{Department of Physics, University of Tokyo, 7-3-1 Hongo, Bunkyo-ku, Tokyo 113-0033, Japan}
  \affiliation{Institute of Integrated Research, Institute of Science Tokyo, 4259 Nagatsuta-cho, Midori-ku, Yokohama, Kanagawa 226-8501, Japan}
\author{Sotatsu~Otabe}%
  \affiliation{Institute of Integrated Research, Institute of Science Tokyo, 4259 Nagatsuta-cho, Midori-ku, Yokohama, Kanagawa 226-8501, Japan}
\author{Tomoya~Sato}%
  \affiliation{Institute of Integrated Research, Institute of Science Tokyo, 4259 Nagatsuta-cho, Midori-ku, Yokohama, Kanagawa 226-8501, Japan}
\author{Martin~Miranda}%
  \affiliation{Institute of Integrated Research, Institute of Science Tokyo, 4259 Nagatsuta-cho, Midori-ku, Yokohama, Kanagawa 226-8501, Japan}
\author{Nobuyuki~Takei}%
  \affiliation{Institute of Integrated Research, Institute of Science Tokyo, 4259 Nagatsuta-cho, Midori-ku, Yokohama, Kanagawa 226-8501, Japan}
\author{Mikio~Kozuma}%
  \affiliation{Institute of Integrated Research, Institute of Science Tokyo, 4259 Nagatsuta-cho, Midori-ku, Yokohama, Kanagawa 226-8501, Japan}
  \affiliation{Department of Physics, Institute of Science Tokyo, 2-13-1 Ookayama, Meguro-ku, Tokyo 152-8550, Japan}

\date{\today}

\begin{abstract}
The sensitivity of an interferometer depends on its readout scheme.
However, little attention has been paid to the readout schemes of atom interferometers from the viewpoint of their sensitivity.
The difference in sensitivity between readout schemes or their optimization has not been considered in the literature.
Herein we analytically calculate the sensitivities of an atom interferometer with typical readout schemes by applying the two-photon formalism, which was developed for optical interferometers to deal with quantum noise.
Our calculations reveal that by using sinusoidal phase modulation, the sensitivity can surpass that obtained by the conventional phase sweeping scheme.
The superiority of this phase modulation scheme for both cold and thermal atomic beams is demonstrated.
In addition, we show that the phase modulation scheme is advantageous for atom-flux fluctuation and resists atom-flux drift.
This study performs a general analysis of the sensitivity of atom interferometers and identifies an advantageous readout scheme.
\end{abstract}

\maketitle


\section{\label{sec:introduction}Introduction}

Light-pulse atom interferometers~\cite{Kasevich:1991zz,Riehle:1991,Kasevich:1992yii,Geiger:2020aeq} have attracted considerable attention as highly sensitive quantum sensors.
They utilize the interactions between atoms and light, such as stimulated Raman transitions or Bragg diffraction, which function as beam splitters or mirrors for atomic beams.
Through a sequence of light pulses, a Mach-Zehnder atom interferometer is formed.
Light-pulse atom interferometers have been used to measure acceleration~\cite{McGuinness:2012,Lautier:2014gaa,Cheiney:2018}, rotation rate~\cite{Gustavson:2000,Durfee:2005fna,Savoie:2018guq,Avinadav:2020}, gravity~\cite{Peters:2001,Bidel:2013oja,Wu:2019ikc}, and gravity gradients~\cite{Snadden:1998zz,Biedermann:2014jya}.
Moreover, they can be applied to fundamental physics because of their high precision.
For example, they can be used to test the weak equivalence principle~\cite{Fray:2004zs,Tarallo:2014oaa,Zhou:2015pna,Rosi:2017ieh} and detect gravitational waves~\cite{Tino:2011zz,Dimopoulos:2007cj,Canuel:2017rrp,MAGIS-100:2021etm}.

In particular, atom interferometers with continuous atomic beams are advantageous because of their high data rates and bandwidths.
In addition, the aliasing noise can be reduced by using continuous atomic beams~\cite{Joyet:2012}.
Historically, after the development of atom interferometers that use thermal atomic beams~\cite{Riehle:1991, Durfee:2005fna, Lenef:1997} (\cite{Durfee:2005fna} is transversely laser cooled), pulsed cold atomic sources were invented~\cite{Geiger:2020aeq}, and continuous cold atomic beams have been demonstrated recently ~\cite{Xue:2015, Kwolek:2020, Kwolek:2021yqp, Meng:2022cbn}.
A thermal atomic beam is taken out of an effusive atomic oven.
The vapor of the atoms continuously flows out of the hole in the oven, allowing for the continuous emission of a thermal atomic beam.
In this paper, a thermal atomic beam refers to an atomic beam without longitudinal laser cooling; for example, an atomic beam emitted from an oven and subjected to transversal cooling is also called a thermal atomic beam.
On the other hand, (longitudinally) laser-cooled atomic beams offer longer interrogation times because they have slower longitudinal velocities than thermal atomic beams.
Consequently, their sensitivities improve as the scale factors increase with the interrogation time.
In addition, their interference fringes show greater contrast and broader dynamic ranges owing to their narrower velocity widths.
In pulsed cold atomic beams, every cycle needs trapping and cooling times; thus, obtaining high data rates is challenging~\cite{Rakholia:2014,Savoie:2018guq}.
However, with the recent progress in experimental techniques, continuous cold atomic beams can be incorporated into atom interferometers~\cite{Xue:2015,Kwolek:2020,Kwolek:2021yqp}.
In addition, the use of thermal atomic beams continues to be active.
Thermal atomic beams are advantageous for sensitivity in that they can achieve much higher flux.
Furthermore, an operation to extend the dynamic range has been invented~\cite{Sato:2023}, whose narrowness was a weak point of thermal atomic beams.
In light of these circumstances, hereafter, we focus on atom interferometers that use continuous atomic beams.

The choice of the readout scheme plays a pivotal role in optimizing the sensitivity of an atom interferometer; the readout scheme defines how the phase signal is derived from the interferometer output.
Sensitivity refers to the noise level converted into a phase-equivalent quantity because phase signals that are buried in noise cannot be detected.
The design sensitivity of an atom interferometer is determined by the shot noise, and the effect of the shot noise on the sensitivity depends on the readout scheme.
Shot noise, also called quantum projection noise~\cite{Itano:1993}, is caused by quantum fluctuation in counting the number of atoms.
Because shot noise arises from its quantum nature, it cannot be suppressed without a quantum manipulation of the state of atoms~\cite{Orzel2001}.
In atom interferometers with continuous atomic beams, shot noise limits their sensitivity after other technical noises are well suppressed.
Therefore, reducing the effect of shot noise is a central issue in atom interferometers.
The readout scheme must be optimized to minimize the effects of the shot noise.

However, little attention has been paid to the readout schemes of atom interferometers from the viewpoint of their sensitivity.
In previous studies~\cite{Gustavson:2000,Kwolek:2021yqp}, the shot-noise level was calculated based on the fact that the counted number of atoms fluctuated with $\sqrt{\mathcal{F}\tau}$, where $\mathcal{F}$ is the average flux and $\tau$ is the integration time.
In the calculation, the atom interferometer was assumed to be set to the midfringe of the interference.
Thus, the output fluctuation was $\sqrt{\mathcal{F}\tau/2}$, the susceptibility of the interferometer from the phase to the output was $\mathcal{F}/2$, and the resulting phase-equivalent power spectral density (PSD) of the shot noise was $4/\mathcal{F}$.
Nevertheless, the phase signal of an atom interferometer is often extracted by linearly sweeping the phase of the matter wave of an atomic beam~\cite{Gustavson:2000,Kwolek:2021yqp}.
In this readout scheme, the phase difference between the two paths in the interferometer evolves in proportion to time.
Consequently, the output of the interferometer oscillates, and the original phase difference can be extracted by a lock-in measurement of the output.
Although this phase sweep scheme has been widely used, the difference between the sensitivity for the phase sweep scheme and that for the static midfringe readout scheme was unknown.
Therefore, the readout scheme has not been optimized for improving the sensitivity of the atom interferometer.

In this paper, we analyze the sensitivity of an atom interferometer by applying the two-photon formalism~\cite{Caves:1985zz, Schumaker:1985zz,Corbitt:2005qv}, which was developed to calculate quantum noise in optical interferometers.
We establish a framework to calculate the sensitivity of an atom interferometer that can handle general readout schemes, including the phase sweep scheme.
Furthermore, we propose using sinusoidal phase modulation for a readout scheme to improve the sensitivity of an atom interferometer.
Our noise analysis reveals that the phase modulation scheme is superior to the phase sweep scheme in terms of sensitivity if a suitable modulation index and fringe point are selected.
We also demonstrate the calculation of atomic beams with broad velocity widths.
The sensitivity of the phase modulation scheme is found to surpass that of the phase sweep scheme, even when the velocity width of an atomic beam is broad.
In addition, our calculation can also deal with atom-flux fluctuations, and we discuss the effect of the atom-flux fluctuations on sensitivity.

\section{\label{sec:noise}Noise analysis with phase modulation}

\subsection{General framework}
\begin{figure}
    \centering
        \includegraphics[width=\columnwidth,clip]{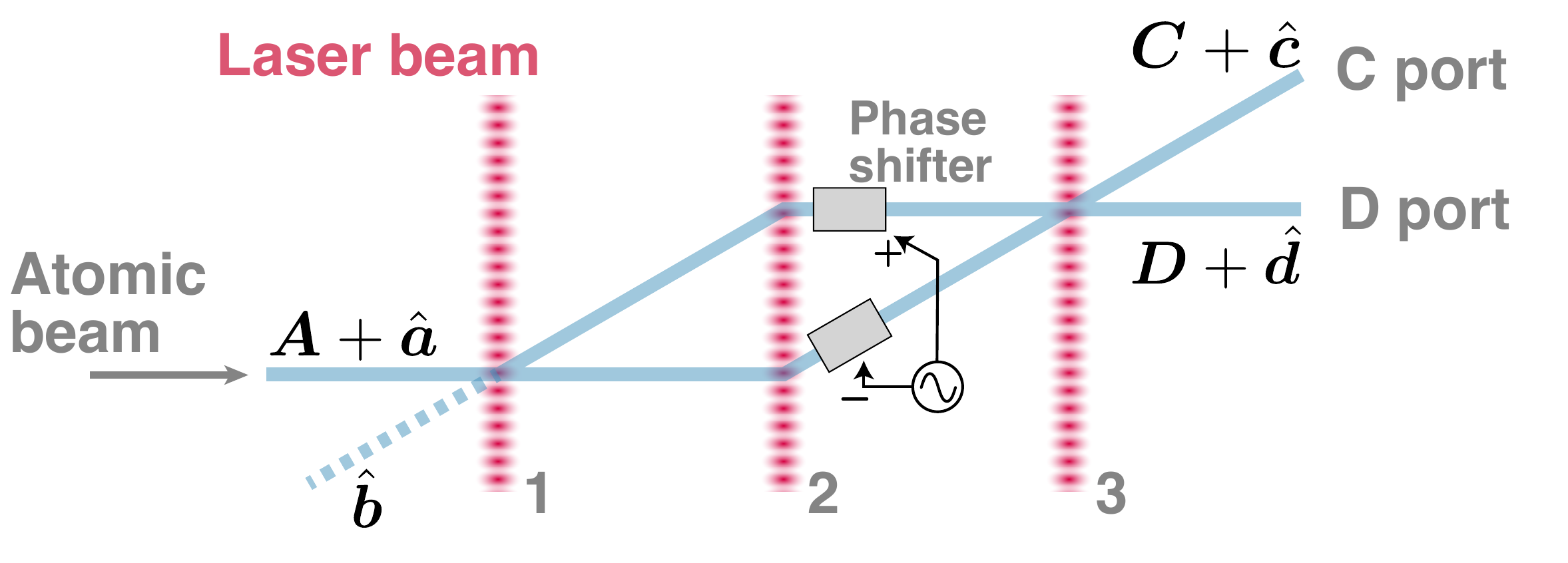}
        \caption{Schematic of an atom interferometer for noise analysis.
        The atomic beam constitutes a Mach-Zehnder interferometer.
        Three pairs of laser beams interact with the atomic beam and function as beam splitters.
        The phase of the atomic beam in each path is differentially modulated.
        In practice, the laser beams also work as the phase shifters, which are described separately for clarity.
        The flux of the atomic beam is measured at the output ports.}
        \label{fig:schematic}
\end{figure}

We consider a typical atom interferometer of the Mach-Zehnder type, as shown in Fig.~\ref{fig:schematic}.
Three pairs of laser beams are applied as beam splitters for the atomic beam.
Let the reflectivity (diffraction efficiency for the atomic beam) and transmissivity of each laser beam be denoted by $r_i$ and $t_i$, respectively, and the subscript $i = 1,2,3$ denote the index of the laser beam.
The laser beams also work as phase modulators; however, we will describe this aspect separately for clarity.
We consider that the phase modulation is differential for two paths, and the atomic beams accept the phase $\pm \theta(t)$ at the modulators.
We note that if the interaction with the laser beam is used as the phase modulators, the period of modulation must be sufficiently larger than the interaction time.

To analyze quantum shot noise in an interferometer, it is necessary to treat the atomic beam within the framework of quantum field theory.
In focusing on the quantum fluctuations in counting the number of atoms, the inner degrees of freedom of the atom can be considered separately.
In addition, the interactions between atoms can be ignored because, in most atom interferometers, the atoms are not densely packed.
In this situation, the field of the atomic beam can be considered to be a scalar field and the field operator is described by
\begin{align}
    \hat{\psi}(z, t) = \int_0^\infty \qty(\hat{a}_\omega e^{\text{i}kz - \text{i} \omega t} + \hat{a}_\omega^\dagger e^{-\text{i}kz + \text{i}\omega t}) \frac{\dd{\omega}}{2\pi},
    \label{fig:field_op}
\end{align}
where $\hat{a}_\omega$ and $\hat{a}_\omega^\dagger$ are the annihilation and creation operators for the mode of the angular frequency $\omega$, respectively, and $k$ is the wave number of the matter wave; the angular frequency and wave number follow the Einstein--de Broglie relation for the matter wave.
Let the propagation direction of the atomic beam be along the $z$ axis.
In our calculation model, we assume that the transverse momentum width of the atomic beam is sufficiently small and can be ignored.
Accordingly, the spatial dependence in the transverse direction is omitted in Eq.~(\ref{fig:field_op}).
The annihilation and creation operators satisfy $[\hat{a}_\omega,~\hat{a}_{\omega'}^\dagger] = 2\pi \delta(\omega - \omega')$ and the commutation relations of the other combinations are zero.
Because an atomic beam field with an atomic beam flux $\mathcal{F}$ has a large steady-state amplitude of $\sqrt{2\mathcal{F}}$, calculations are facilitated by using the quadrature basis around the angular frequency of the matter wave of the atomic beam.
Thus, the field operator can be rewritten as~\cite{Caves:1985zz,Schumaker:1985zz}
\begin{align}
    \hat{\psi}(t) = \left[\sqrt{2\mathcal{F}} + \hat{a}_1(t)\right] \cos\omega_0t + \hat{a}_2(t) \sin\omega_0t,
\end{align}
where $\omega_0$ denotes the angular frequency of the matter wave, $\hat{a}_1(t)$ denotes the amplitude quadrature, and $\hat{a}_2(t)$ denotes the phase quadrature.
We describe the quadratures of the fluctuations as vectors $\hat{\bm{a}} = (\hat{a}_1~\hat{a}_2)\!^\top$ and denote the carrier field as $\bm{A} = (\sqrt{2\mathcal{F}}~0)\!^\top$.
As shown in Fig.~\ref{fig:schematic}, we also define the incoming quantum vacuum field from the other input port as $\hat{\bm{b}} = (\hat{b}_1~\hat{b}_2)\!^\top$, the output fields containing fluctuations as $\hat{\bm{c}}$ and $\hat{\bm{d}}$, and the modulated output carrier fields as $\bm{C}$ and $\bm{D}$.

By introducing phase modulation, we obtain the output of the fluctuation in the form of $\bm{P}(t)\vdot \hat{\bm{a}} + \bm{Q}(t)\vdot \hat{\bm{b}}$ after calculating the input-output relation of the interferometer.
The modulation to the carrier is represented by the coefficients $\bm{P}(t)$ and $\bm{Q}(t)$.
Then, it was shown in the context of optical systems~\cite{Buonanno:2003ch} that the PSD of the fluctuation $S$ is expressed by
\begin{align}
    S = \frac{1}{T}\int_0^T D^2(t)\qty[\norm{\bm{P}(t)}^2 + \norm{\bm{Q}(t)}^2] \dd{t},
    \label{eq:yanbei}
\end{align}
where $D(t)$ is the demodulation function, $T$ is the period of the demodulation function, and $\norm{\cdot}$ is the norm of the vector.
This PSD is calculated based on the fact that the PSD of the vacuum-field fluctuation, such as $\hat{\bm{a}}$ or $\hat{\bm{b}}$, is $1$.
During the derivation of Eq.~(\ref{eq:yanbei}), we only assume that the fluctuation $\hat{\bm{a}}$ in the atomic beam is as large as that in the vacuum field.
In other words, our assumption is conventional, in that the counted number of atoms fluctuates with the square root of the average number.

For our system, we can calculate the input-output relation of the fields as
\begin{align}
  \begin{split}
    \hat{\bm{c}} &= \left[r_1 r_2 r_3 R(\theta(t)) + t_1 r_2 t_3 R(-\theta(t))\right] \hat{\bm{a}} \\
            & \qquad + \left[t_1 r_2 r_3 R(\theta(t)) - r_1 r_2 t_3 R(-\theta(t))\right] \hat{\bm{b}},
    \end{split} \label{eq:cport}\\
  \begin{split}
    \hat{\bm{d}} &= \left[r_1 r_2 t_3 R(\theta(t)) - t_1 r_2 r_3 R(-\theta(t))\right] \hat{\bm{a}} \\
            & \qquad + \left[t_1 r_2 t_3 R(\theta(t)) + r_1 r_2 r_3 R(-\theta(t))\right] \hat{\bm{b}},
    \end{split} \label{eq:dport}
\end{align}
where we define the rotation matrix as
\begin{align}
    R(x) =
      \left(
          \begin{array}{cc}
            \cos x & -\sin x \\
            \sin x & \cos x
          \end{array}
      \right).
\end{align}
In a later section, we discuss the case in which the atomic beam has a broad velocity distribution.
Here we begin with the case in which the atomic beam has a fixed velocity.
Then the reflectivities and transmissivities are constant for all atoms; thus, let us assume $r_1 = t_1 = r_3 = t_3 = 1/\sqrt{2}$ and $r_2 = 1$.
We note that the atomic beam velocity is different from the phase velocity of the matter wave and it affects shot noise only via reflectivities and transmissivities.
Because the output ports C and D are symmetric and result in equivalent noise calculations, we focus on port C.
Then the input-output relation is simplified as
\begin{align}
  \hat{\bm{c}} &=
    \left(
        \begin{array}{cc}
          \cos\theta(t) & 0 \\
          0 & \cos\theta(t)
        \end{array}
    \right)
    \hat{\bm{a}} + 
    \left(
        \begin{array}{cc}
          0 & -\sin\theta(t)\\
          \sin\theta(t) & 0
        \end{array}
    \right)
    \hat{\bm{b}}.
\end{align}
The modulated carrier at the output port C can be calculated as
\begin{align}
    \bm{C} = \cos\theta(t) \mqty(\sqrt{2\mathcal{F}} \\ 0).
\end{align}
Therefore, the shot noise in the output flux $\delta \hat{\mathcal{F}}^\text{shot} = \bm{C} \vdot \hat{\bm{c}}$ is expressed as:
\begin{align}
    \delta \hat{\mathcal{F}}^\text{shot} = \frac{\sqrt{2\mathcal{F}}}{2}\qty(\qty{1 + \cos\qty[2\theta(t)]}\hat{a}_1 - \sin\qty[2\theta(t)]\hat{b}_2),
    \label{eq:flux_shotnoise}
\end{align}
while the second-order terms of the fluctuations are ignored.
Equation~(\ref{eq:flux_shotnoise}) implies that
\begin{align}
    \bm{P}(t) &= \mqty(\sqrt{2\mathcal{F}}\qty{1 + \cos\qty[2\theta(t)]}/2 \\ 0), \\
    \bm{Q}(t) &= \mqty(0 \\ -\sqrt{2\mathcal{F}}\sin\qty[2\theta(t)]/2)
\end{align}
for our model.
Thus, the PSD of shot noise is calculated using Eq.~(\ref{eq:yanbei}) as follows:
\begin{align}
    S^\text{shot}_\mathcal{F} = \frac{\mathcal{F}}{2} \frac{1}{T} \int_0^T D^2(t) \qty(\qty{1 + \cos\qty[2\theta(t)]}^2 + \sin\qty[2\theta(t)]^2) \dd{t}.
    \label{eq:shotnoise}
\end{align}
To evaluate the sensitivity, the PSD of the shot noise is converted into the PSD of the phase-equivalent shot noise as $S^\text{shot}_{\phi} = S^\text{shot}_\mathcal{F} / \chi^2$, where $\chi$ is the susceptibility of the interferometer. 
The susceptibility is defined by the conversion coefficient from the phase difference $\phi_\text{s}$ in the atomic beams between the two paths in the interferometer to the output signal.
The phase $\phi_\text{s}$ can be incorporated in the above calculations as the offset in the modulation phase.
Therefore, the output flux $\mathcal{F}_\text{out}$ including $\phi_\text{s} (\ll 1)$ is expressed by
\begin{align}
    \mathcal{F}_\text{out} = \frac{1}{2}\norm{\bm{C}}^2 = \frac{\mathcal{F}}{2}\qty{1 + \cos\qty[2\theta(t)] + \phi_\text{s}\sin\qty[2\theta(t)]}.
\end{align}
As the derivative with respect to the phase difference is $\dv{\mathcal{F}_\text{out}}{\phi_\text{s}} = \qty{\mathcal{F}\sin[2\theta(t)]}/2$, its demodulated signal is $\chi$.
We do not consider the sign of susceptibility $\chi$ in the calculation of sensitivity; thus, we will use its absolute value.

\subsection{The phase sweep scheme and the phase modulation scheme}
Before considering specific forms of modulation, we confirm that our general expression for shot noise in Eq.~(\ref{eq:shotnoise}) is consistent with the shot-noise levels assumed in previous studies ~\cite{Gustavson:2000,Kwolek:2021yqp}.
For the midfringe readout scheme without modulation, we can describe that $\theta(t) = \pi/4$, $D(t) = 1$, and $\chi = \mathcal{F}/2$.
Thus, from Eq.~(\ref{eq:shotnoise}), the phase-equivalent shot noise is calculated as $4/\mathcal{F}$, which is consistent with the values of previous studies.
In practice, the phase sweep scheme has often been used in experiments with atom interferometers.
In the phase sweep method, $\theta(t) = \omega_\text{s}t$ and $D(t) = \sin2\omega_\text{s}t$, where $\omega_\text{s}$ is the angular frequency of the sweep.
The calculated PSD of the shot noise is $8/\mathcal{F}$, and a shot noise level of $4/\mathcal{F}$ is no longer achieved with the phase sweep scheme.

\begin{figure}
    \centering
        \includegraphics[width=\columnwidth,clip]{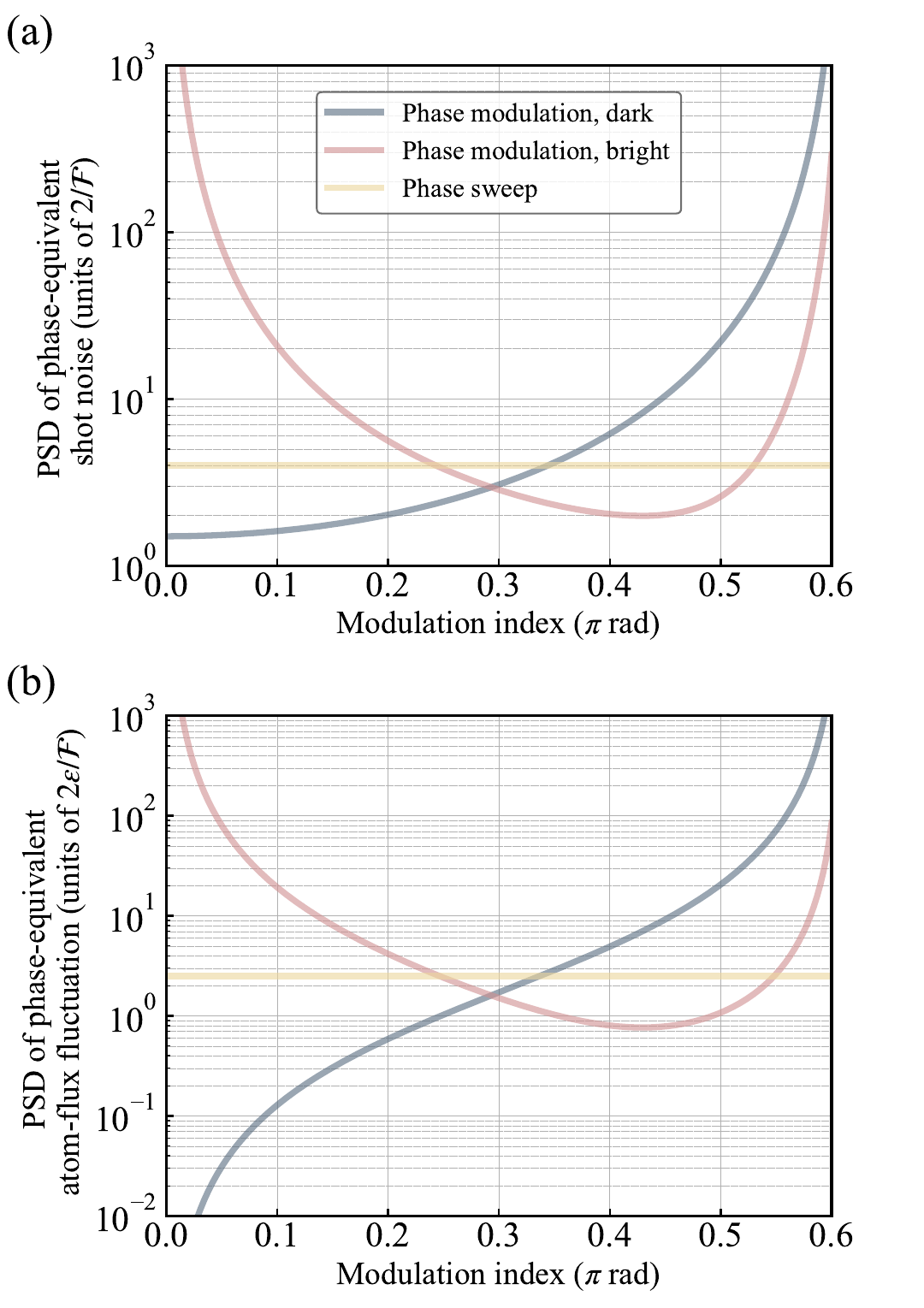}
        \caption{Power spectral density of phase-equivalent shot noise and atom-flux fluctuation according to the modulation index.
        In this plot, a fixed velocity is assumed for all atoms.
        For the phase modulation scheme, the noises are calculated at both the dark and bright fringes.
        As the phase sweep scheme does not have the parameter of a modulation index, its plot is shown by a flat line.}
        \label{fig:psd_cold}
\end{figure}

To improve the sensitivity, we propose the use of sinusoidal modulation instead of the phase sweep scheme.
In other words, we introduce phase modulation $\theta(t) = \phi_\text{m}\sin\omega_\text{m}t$ and demodulation $D(t) = \sin\omega_\text{m}t$, where $\omega_\text{m}$ is the modulation angular frequency and $\phi_\text{m}$ is the modulation index.
For phase modulation, the susceptibility of the interferometer is maximal at the dark and bright fringes.
For the output port C, the default fringe is bright and the dark fringe can be chosen by adding a phase offset as $\theta(t) = \phi_\text{m}\sin\omega_\text{m}t + \pi/2$.
The calculated susceptibilities and PSDs of phase-equivalent shot noise are summarized in Table~\ref{tab:cold} and plotted in Fig.~\ref{fig:psd_cold}(a), where $J_n(x)$ is the Bessel function of the first kind.
The best sensitivity is obtained using the phase modulation scheme at the dark fringe with a sufficiently small modulation index, as $S^\text{shot}_\phi = 3/\mathcal{F}$.
At the bright fringe, the sensitivity exceeds that of the phase sweep scheme and is optimal with $\phi_\text{m} \simeq 0.43\pi$; however, it does not reach that of the dark fringe.
Although, in general, a small modulation index is preferable for the dark fringe, it should be noted that a small modulation index makes the interferometer more sensitive to other noises on the output port.
A small modulation index leads to small susceptibility.
Thus, the phase signal is small at the output port, though the shot noise is also small.
Consequently, the sensitivity will deteriorate when the signal is buried by other noises.
Therefore, the modulation index should be sufficient so that the modulation is not covered by other noises.

While we can intuitively expect the different results between the dark and bright fringes where the modulation index is small, the whole behavior is elucidated by the analytical calculations.
In comparing the dark and bright fringes, the susceptibility is the same.
On the other hand, if the modulation index is small, the bright fringe has a large carrier, which introduces extra noise.
Consequently, the sensitivity is poor for the bright fringe.
However, the behavior has been unpredictable where the modulation index is large because infinite sidebands, represented by an infinite series of Bessel functions, contribute to the signal and noise.
The analytical calculations clarify the whole behavior, and the calculated results shown in Table~\ref{tab:cold} can be expressed in the form where the infinite series is reduced.

\begin{table*}
    \caption{Analytical expressions of the susceptibilities and power spectral densities of phase-equivalent shot noise and classical atom-flux fluctuation.}
    \centering
    \begin{ruledtabular}
    \begin{tabular}{l l l l}
      Readout scheme & Susceptibility & Shot noise & Atom-flux fluctuation \rule[-2.5mm]{0mm}{6mm} \\
      \colrule
      \makecell[l]{Phase modulation \\ at the dark fringe} & $\begin{aligned} \frac{\mathcal{F}}{2}J_1(2\phi_\text{m}) \end{aligned}$
              & $\begin{aligned} \frac{2}{\mathcal{F}}\frac{1 - J_0(2\phi_\text{m}) + J_2(2\phi_\text{m})}{[J_1(2\phi_\text{m})]^2} \end{aligned}$
              & $\begin{aligned} \frac{2\varepsilon}{\mathcal{F}}\frac{1 - J_0(2\phi_\text{m}) + J_2(2\phi_\text{m}) - \frac{1}{4}\qty[1 - J_0(4\phi_\text{m}) + J_2(4\phi_\text{m})]}{[J_1(2\phi_\text{m})]^2} \end{aligned}$ \vspace{1em} \rule[0mm]{0mm}{6mm} \\
      \makecell[l]{Phase modulation \\ at the bright fringe} & $\begin{aligned} \frac{\mathcal{F}}{2}J_1(2\phi_\text{m}) \end{aligned}$
               & $\begin{aligned} \frac{2}{\mathcal{F}}\frac{1 + J_0(2\phi_\text{m}) - J_2(2\phi_\text{m})}{[J_1(2\phi_\text{m})]^2} \end{aligned}$
               & $\begin{aligned} \frac{2\varepsilon}{\mathcal{F}}\frac{1 + J_0(2\phi_\text{m}) - J_2(2\phi_\text{m}) - \frac{1}{4}\qty[1 - J_0(4\phi_\text{m}) + J_2(4\phi_\text{m})]}{[J_1(2\phi_\text{m})]^2} \end{aligned}$ \vspace{1em} \\
      Phase sweep & $\begin{aligned} \frac{\mathcal{F}}{4} \end{aligned}$
         & $\begin{aligned} 4\times \frac{2}{\mathcal{F}} \end{aligned}$
         & $\begin{aligned} \frac{5}{2}\times \frac{2\varepsilon}{\mathcal{F}} \end{aligned}$ \rule[-3mm]{0mm}{0mm} \\
    \end{tabular}
    \end{ruledtabular}
    \label{tab:cold}
\end{table*}

Our calculation can also be applied to classical atom-flux fluctuation by setting $\hat{a}_2 = \hat{b}_1 = \hat{b}_2 = 0$ because there is no vacuum fluctuation classically.
As the level of the atom-flux fluctuation is arbitrary, i.e., dependent on the quality of the atomic beam, let us suppose that the power of the atom-flux fluctuation is $\varepsilon$ times greater than that of the vacuum fluctuation ($\hat{a}_1 \rightarrow \sqrt{\varepsilon} \hat{a}_1$).
The results for the atom-flux fluctuations are listed in Table~\ref{tab:cold} and plotted in Fig.~\ref{fig:psd_cold}(b).
Notably, the atom-flux fluctuation can be canceled completely in the phase modulation scheme at the dark fringe.

\subsection{Atomic beam with a finite velocity width \label{sec:thermal}}
In general, an atomic beam is not always ideally cooled, or even cooled.
Therefore, we consider an atom interferometer using an atomic beam with a finite velocity width.
Atoms with different velocities exhibit different interaction times with light.
Consequently, the reflectivities and transmissivities of the interactions vary and depend on velocity $v$; thus, $r_1 = r_3 = r(v)$, $t_1 = t_3 = \sqrt{1 - r^2(v)}$, $r_2 = r'(v)$, and $t_2 = \sqrt{1 - r'^2(v)}$.
Let us assume that the three pairs of laser beams are adjusted to a certain velocity $v_0$ to achieve $r(v_0) = 1/\sqrt{2}$ and $r'(v_0) = 1$.
As the interaction time is inversely proportional to the velocity and the atoms undergo Rabi oscillations, the reflectivities are written as $r(v) = \sin\qty[(\pi/4)(v_0/v)]$ and $r'(v) = \sin\qty[(\pi/2)(v_0/v)]$.
It is straightforward to calculate the PSD of the shot noise $S_\mathcal{F}(v)$ and susceptibility $\chi(v)$.
See Tables~\ref{tab:thermal_shot} and~\ref{tab:thermal_intensity} in Appendix~\ref{sec:thermal_analytical} for the analytical expressions of the results, which include the case of atom-flux fluctuation.
In the calculations, the fields that leak at the second laser beam are also taken into account because $t_2$ is not zero.
Because it is difficult to spatially isolate the leakage fields, they are observed at the output port simultaneously after the third interaction with the laser beam, which causes additional noise.
Note that, in contrast to the fixed velocity case, the results differ between ports C and D, as can be inferred from Eqs.~(\ref{eq:cport}) and~(\ref{eq:dport}).
The difference arises when the reflectivity and transmissivity are different.
The atomic beam at port C results from the interference of the atomic beams either both reflected or both transmitted by the first and the third laser beams.
In contrast, the atomic beam at port D arises from the interference between the atomic beams transmitted once and reflected once by the first and third laser beams.
Consequently, if the reflectivity and transmissivity cannot be set to $1/\sqrt{2}$ for all atoms, differences will arise between ports C and D.
Thus, the calculations for both ports are presented herein.
Finally, we obtain the PSD of the phase-equivalent shot noise $S^\text{shot}_\phi$ as
\begin{align}
    S^\text{shot}_\phi =  \frac{\int_0^\infty p(v) S^\text{shot}_\mathcal{F}(v)\dd{v}}{\qty[\int_0^\infty p(v) \chi(v)\dd{v}]^2},
\end{align}
where $p(v)$ is the velocity distribution of the atoms.

\begin{figure}
    \centering
        \includegraphics[width=\columnwidth,clip]{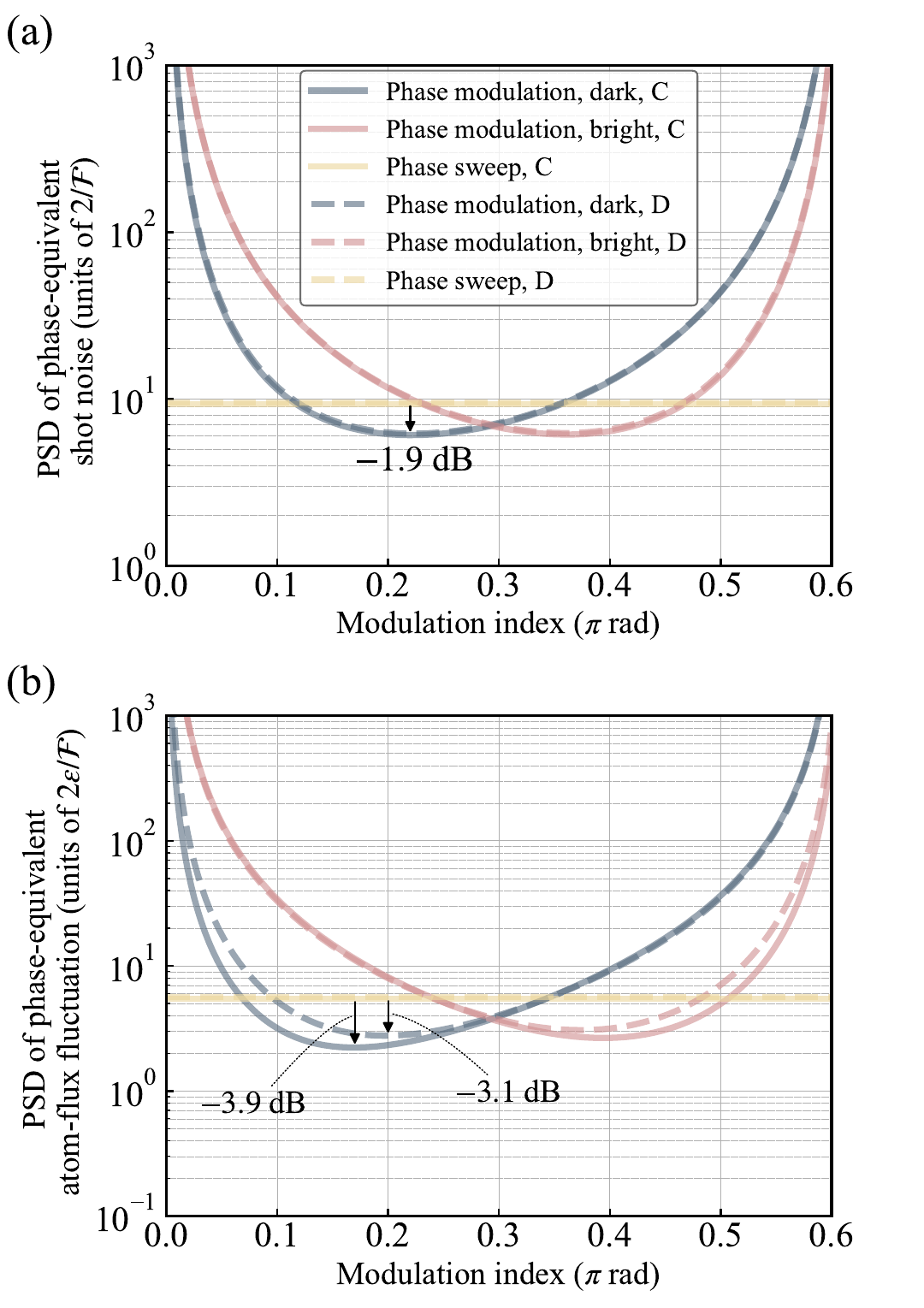}
        \caption{Power spectral density of phase-equivalent shot noise and atom-flux fluctuation in the case of a thermal atomic beam.}
        \label{fig:psd_thermal}
\end{figure}

As an essential example, we present the calculated results for a thermal atomic beam.
A thermal atomic beam is introduced from the atomic vapor in the oven through a small hole, and it propagates without any laser cooling.
Assuming that the vapor is in thermal equilibrium, the velocity distribution of the thermal atomic beam is described by~\cite{Ramsey:1986}
\begin{align}
    p(v) = 2\qty(\frac{m}{2k_\text{B}T})^2 v^3\exp\qty(-\frac{mv^2}{2k_\text{B}T}),
\end{align}
where $k_\text{B}$ is the Boltzmann constant, $T$ is the temperature, and $m$ is the mass of the atom.
Here we assume that the laser beams are tuned for the atoms with the most probable speed, specifically, $v_0 = \sqrt{3k_\text{B}T/m}$.
The results are presented in Fig.~\ref{fig:psd_thermal}.
Notably, the results are independent of the temperature and mass of the atoms.
For a thermal atomic beam, the contrast in the interference is reduced because of the finite velocity width.
Consequently, in regions where the modulation index is small and the susceptibility is also small, the effect of noise from background atoms that do not contribute to interference increases.
Therefore, the sensitivity is poor when the modulation index is small.
For the dark fringe, the modulation index is optimal at approximately $0.2\pi$.
For shot noise, the PSD values under the phase modulation scheme are up to 0.65 times smaller (\SI{-1.9}{dB}) than those under the phase sweep scheme for both ports C and D.
These sensitivities can be obtained at either the dark or bright fringes by selecting a suitable modulation index.
In contrast, dark fringes are favorable in the case of atom-flux fluctuations.
The PSD value under the phase modulation scheme is 0.41 times smaller (\SI{-3.9}{dB}) at port C and 0.49 times smaller (\SI{-3.1}{dB}) at port D than that under the phase sweep scheme.
In any case, the phase modulation scheme is advantageous with optimal modulation indices.

A thermal atomic beam is an example of an atomic beam with a wide velocity distribution.
For realistic laser-cooled atomic beams, we note that the atomic beams produced by state-of-the-art laser cooling techniques have sufficiently narrow velocity widths that can be adequately modeled using our calculation without considering the velocity width.
See Appendix~\ref{sec:cooled} for an example of a laser-cooled atomic beam with the recently reported experimental parameters.

\section{\label{sec:discussions}Discussions}
\subsection{Resistance to the atom-flux drift}
In the previous sections we considered atom interferometer sensitivities, which included the effect of stationary noise.
Because the stability of measurement for an atom interferometer is also crucial in practical applications, we discuss ways to mitigate the effect of atom-flux drift for both readout schemes.
Although we focus on atom-flux drift, our discussion below applies to any drift that scales the overall intensity of the interferometer output.
When the atom-flux drift scales the output, the output deviation is indistinguishable from the phase signal.
However, in general, the normalization of the output suppresses the effect of atom-flux drift.

In the phase sweep scheme, dual-phase lock-in detection can suppress the effect of the drift without sacrificing sensitivity.
The output signal oscillates at twice the sweep frequency in the phase sweep scheme.
Thus, dual-phase lock-in detection can determine the phase by taking the ratio of the two orthogonal quadrature components instead of demodulating using a single demodulation function.
By taking this ratio, the overall factor of the output does not influence the determined phase; therefore, the effect of atom-flux drift is avoided.
We note that the sensitivity is independent of whether demodulation is achieved by a single demodulation function or by using dual-phase lock-in detection.

The effect of atom-flux drift can also be removed in the phase modulation scheme using a signal demodulated at twice the modulation frequency, as proposed for fiber-optic gyroscopes~\cite{Bohm:1983}.
The phase difference in the interferometer $\phi_\text{s}$ can be expressed as
\begin{align}
    \phi_\text{s} = \tan^{-1}\qty(\frac{J_2\qty(\phi_\text{m})}{J_1\qty(\phi_\text{m})}\frac{s_1}{s_2}),
\end{align}
where $s_1$ and $s_2$ are the amplitudes of the first- and second-harmonic components in the output, respectively, which are extracted via demodulation at $\omega_\text{m}$ and $2\omega_\text{m}$.
As in the phase sweep scheme, the overall scale change in the output due to the atom-flux drift is negated in the fractions of $s_1$ and $s_2$.
Therefore, in terms of stability, both the phase sweep and phase modulation schemes can resist atom-flux drift.

\subsection{Phase modulation and demodulation with a square wave}
We have studied a phase modulation readout scheme with a sinusoidal function.
Here we note the possibility of further improving the sensitivity by optimizing the waveforms of the modulation and demodulation functions.
In a certain case, it is known that optical interferometers with square-wave modulation and demodulation have higher sensitivity~\cite{Miranda:2023}.
However, only the case of a dark fringe with a sufficiently small modulation index has been discussed previously~\cite{Somiya:2002tx,Buonanno:2003ch}.
In this case, the combination of square-wave modulation and square-wave demodulation resulted in a better shot-noise level, which is $2/3$ the level of the PSD for sinusoidal modulation and demodulation.

However, unlike optical interferometers, atomic beams can only be modulated relatively slowly.
This is because of the longer interaction time between the atomic beam and the modulating laser beam.
This imposes a limitation on the modulation waveform; square-wave modulation cannot be ideally realized because a square wave includes harmonics.
On the other hand, greater flexibility is allowed for the demodulation waveform.
In a sinusoidal phase modulation scheme with square-wave demodulation, the shot noise can be $8/\pi^2 \simeq 0.81$ times lower than that obtained with sinusoidal demodulation.
In contrast, a PSD of the phase-equivalent shot noise of $\pi^2/\mathcal{F}$ is obtained in the phase sweep scheme with square-wave demodulation.
This value is larger than the sinusoidal modulation value of $8/\mathcal{F}$.
The sensitivity of the phase sweep scheme cannot be optimized using square-wave demodulation.
As implied in this case, the general optimization of the demodulation function is not trivial, particularly for thermal atomic beams.
Therefore, it is worthwhile to further explore the optimization of the demodulation function for future work.

\section{\label{sec:conclusion}Conclusion}
We have calculated the sensitivity of atom interferometers with nonstationary carrier fields and with this analysis we have elucidated the usefulness of phase modulation in a readout scheme.
Despite the considerable interest in atom interferometers with continuous atomic beams as highly sensitive sensors, their sensitivities for each readout scheme were not fully understood.
In this work we performed an analysis to reveal the differences in sensitivities of different readout schemes.
We identified that the conventionally assumed sensitivity cannot be achieved with the phase sweep scheme, although it is widely used.
We proposed the use of a phase modulation scheme to obtain higher sensitivity and demonstrated that the phase modulation scheme with the optimal modulation index is superior to the phase sweep and midfringe schemes in terms of sensitivity, even when the atomic beam has a finite velocity width.
Furthermore, we found that the phase modulation scheme also has advantages over atom-flux fluctuation and provides a means to resist atom-flux drift.

Our work will help in the future design of atom interferometers.
Using our analysis, the sensitivities of conventional and new readout schemes can be calculated and compared to select a suitable readout scheme.
While we have focused on sinusoidal waveform modulation, it merits further study to examine more general forms of readout schemes, such as square-wave demodulation.
It would also be worthwhile to explore a colored noise spectrum to address noise sources other than white noises examined in this work.

\begin{acknowledgments}
We thank Takashi~Mukaiyama for helpful discussions.
This work was supported by the Japan Science and Technology Agency under Grants JPMJMI17A3 and JPMJPF2015.
\end{acknowledgments}

\appendix
\section{\label{sec:thermal_analytical} Analytical expressions of our calculation for an arbitrary atomic beam velocity.}
In Sec.~\ref{sec:thermal} we calculated the susceptibilities and PSDs of shot noise and atom-flux fluctuation for an arbitrary atomic beam velocity.
Here we provide the calculated analytical expressions in Tables~\ref{tab:thermal_shot} and~\ref{tab:thermal_intensity}.

\begin{table*}
    \caption{Analytical expressions of the susceptibilities and power spectral densities of shot noise for an arbitrary atomic beam velocity.}
    \centering
    \begin{ruledtabular}
    \begin{tabular}{l l l}
      Readout scheme & Susceptibility & Shot noise \rule[-2.5mm]{0mm}{6mm} \\
      \colrule
      C port & & \rule[0mm]{0mm}{4mm} \\
      \quad Phase modulation at the dark fringe & $2r^2(1 - r^2)r'^2 J_1(2\phi_\text{m})\mathcal{F}$
              & $\begin{aligned}
                   &\mathcal{F}\left\{r'^2 - 2r^2\qty(1 - r^2)\qty(-1 + 2r'^2) \right. \\
                   &\qquad \left. + 2r^2(1 - r^2)r'^2\qty[2J_2(2\phi_\text{m}) - J_1(2\phi_\text{m})/\phi_\text{m}]\right\}
                  \end{aligned}$ \vspace{1em} \\
      \quad Phase modulation at the bright fringe & $2r^2(1 - r^2)r'^2 J_1(2\phi_\text{m})\mathcal{F}$
                & $\begin{aligned}
                    &\mathcal{F}\left\{r'^2 - 2r^2\qty(1 - r^2)\qty(-1 + 2r'^2) \right. \\
                    &\qquad \left. - 2r^2(1 - r^2)r'^2\qty[2J_2(2\phi_\text{m}) - J_1(2\phi_\text{m})/\phi_\text{m}]\right\}
                   \end{aligned}$ \vspace{1em} \\
      \quad Phase sweep & $r^2(1 - r^2)r'^2\mathcal{F}$
         & $\mathcal{F}\qty[2r^2 - 2r^4 + \qty(1 - 2r^2)^2r'^2]$ \vspace{0.2em} \rule[-3mm]{0mm}{0mm} \\

      \colrule
      D port & & \rule[0mm]{0mm}{4mm} \\
      \quad Phase modulation at the dark fringe & $2r^2(1 - r^2)r'^2 J_1(2\phi_\text{m})\mathcal{F}$
              & $\begin{aligned}
                  &\mathcal{F}\left\{1 - r'^2 + 2r^2\qty(1 - r^2)\qty(-1 + 2r'^2) \right. \\
                  &\qquad \left. - 2r^2\qty(1 - r^2)r'^2\qty[J_1(2\phi_\text{m})/\phi_\text{m} - 2J_2(2\phi_\text{m})]\right\}
                 \end{aligned}$ \vspace{1em} \\
      \quad Phase modulation at the bright fringe & $2r^2(1 - r^2)r'^2 J_1(2\phi_\text{m})\mathcal{F}$
              & $\begin{aligned}
                  &\mathcal{F}\left\{1 - r'^2 + 2r^2\qty(1 - r^2)\qty(-1 + 2r'^2) \right. \\
                  &\qquad \left. + 2r^2\qty(1 - r^2)r'^2\qty[J_1(2\phi_\text{m})/\phi_\text{m} - 2J_2(2\phi_\text{m})]\right\}
                 \end{aligned}$ \vspace{1em} \\
      \quad Phase sweep & $r^2(1 - r^2)r'^2\mathcal{F}$
         & $\mathcal{F}\qty[1 - 2r^2 + 2r^4 - \qty(1 - 2r^2)^2r'^2]$ \rule[-3mm]{0mm}{0mm} \\
    \end{tabular}
    \end{ruledtabular}
    \label{tab:thermal_shot}
\end{table*}
\begin{table*}
    \caption{Analytical expressions of the power spectral densities of atom-flux fluctuation for an arbitrary atomic beam velocity.}
    \centering
    \begin{ruledtabular}
    \begin{tabular}{l l}
      Readout scheme & Atom-flux fluctuation \rule[-2.5mm]{0mm}{6mm} \\
      \colrule
      C port & \rule[0mm]{0mm}{4mm} \\
      \quad \makecell[l]{Phase modulation \\ at the dark fringe}
             & $\begin{aligned}
                 &\varepsilon \mathcal{F}\left(4r^4\qty(1 - r^2)^2 + 4r^2\qty(1 - 2r^2)^2\qty(1 - r^2)r'^2 + \qty(1 - 8r^2 + 26r^4 - 36r^6 + 18r^8)r'^4 \right. \\
                 &\qquad - r^2\qty(1 - r^2)r'^2\left\{\qty[r'^2 - 2r^2\qty(1 - r^2)\qty(-1 + 2r'^2)]\qty[4J_1(2\phi_\text{m})/\phi_\text{m} - 8J_2(2\phi_\text{m})] \right. \\
                 &\qquad \qquad \qquad \qquad \qquad \left. \left. + r^2\qty(1 - r^2)r'^2\qty[-J_1(4\phi_\text{m})/\phi_\text{m} + 4J_2(4\phi_\text{m})]\right\} \right)
                \end{aligned}$ \vspace{1em} \\
      \quad \makecell[l]{Phase modulation \\ at the bright fringe}
             & $\begin{aligned}
                 &\varepsilon \mathcal{F}\left(4r^4\qty(1 - r^2)^2 + 4r^2\qty(1 - 2r^2)^2\qty(1 - r^2)r'^2 + \qty(1 - 8r^2 + 26r^4 - 36r^6 + 18r^8)r'^4 \right. \\
                 &\qquad - r^2\qty(1 - r^2)r'^2\left\{-\qty[r'^2 - 2r^2\qty(1 - r^2)\qty(-1 + 2r'^2)]\qty[4J_1(2\phi_\text{m})/\phi_\text{m} - 8J_2(2\phi_\text{m})] \right. \\
                 &\qquad \qquad \qquad \qquad \qquad \left. \left. + r^2\qty(1 - r^2)r'^2\qty[-J_1(4\phi_\text{m})/\phi_\text{m} + 4J_2(4\phi_\text{m})]\right\} \right)
                \end{aligned}$ \vspace{1em} \\
      \quad Phase sweep & $\varepsilon \mathcal{F} \qty[4r^4\qty(1 - r^2)^2 + 4r^2\qty(1 - 2r^2)^2\qty(1 - r^2)r'^2 + \qty(1 - 8r^2 + 25r^4 - 34r^6 + 17r^8)r'^4]$ \vspace{0.2em} \rule[-3mm]{0mm}{0mm} \\

      \colrule
      D port & \rule[0mm]{0mm}{4mm} \\
      \quad \makecell[l]{Phase modulation \\ at the dark fringe}
             & $\begin{aligned}
                 &\varepsilon \mathcal{F}\left(\qty[1 - 2r^2 + 2r^4]^2 - 2\qty(1 - 2r^2)^2\qty(1 - 2r^2 + 2r^4)r'^2 + \qty(1 - 8r^2 + 26r^4 - 36r^6 + 18r^8)r'^4 \right. \\
                 &\qquad - r^2\qty(1 - r^2)r'^2\left\{-\qty[-1 + r'^2 - 2r^2\qty(1 - r^2)\qty(-1 + 2r'^2)]\qty[4J_1(2\phi_\text{m})/\phi_\text{m} - 8J_2(2\phi_\text{m})] \right. \\
                 &\qquad \qquad \qquad \qquad \qquad \left. \left. + r^2\qty(1 - r^2)r'^2\qty[-J_1(4\phi_\text{m})/\phi_\text{m} + 4J_2(4\phi_\text{m})] \right\} \right)
                \end{aligned}$ \vspace{1em} \\
      \quad \makecell[l]{Phase modulation \\ at the bright fringe}
             & $\begin{aligned}
                 &\varepsilon \mathcal{F}\left(\qty[1 - 2r^2 + 2r^4]^2 - 2\qty(1 - 2r^2)^2\qty(1 - 2r^2 + 2r^4)r'^2 + \qty(1 - 8r^2 + 26r^4 - 36r^6 + 18r^8)r'^4 \right. \\
                 &\qquad - r^2\qty(1 - r^2)r'^2\left\{\qty[-1 + r'^2 - 2r^2\qty(1 - r^2)\qty(-1 + 2r'^2)]\qty[4J_1(2\phi_\text{m})/\phi_\text{m} - 8J_2(2\phi_\text{m})] \right. \\
                 &\qquad \qquad \qquad \qquad \qquad \left. \left. + r^2\qty(1 - r^2)r'^2\qty[-J_1(4\phi_\text{m})/\phi_\text{m} + 4J_2(4\phi_\text{m})] \right\} \right)
                \end{aligned}$ \vspace{1em} \\
      \quad Phase sweep & $\varepsilon \mathcal{F} \qty[\qty(1 - 2r^2 + 2r^4)^2 - 2\qty(1 - 2r^2)^2\qty(1 - 2r^2 + 2r^4)r'^2 + \qty(1 - 8r^2 + 25r^4 - 34r^6 + 17r^8)r'^4]$ \rule[-3mm]{0mm}{0mm}
    \end{tabular}
    \end{ruledtabular}
    \label{tab:thermal_intensity}
\end{table*}

\section{\label{sec:cooled} Sensitivity with a cooled atomic beam}
\begin{figure}[b]
    \centering
        \includegraphics[width=\columnwidth,clip]{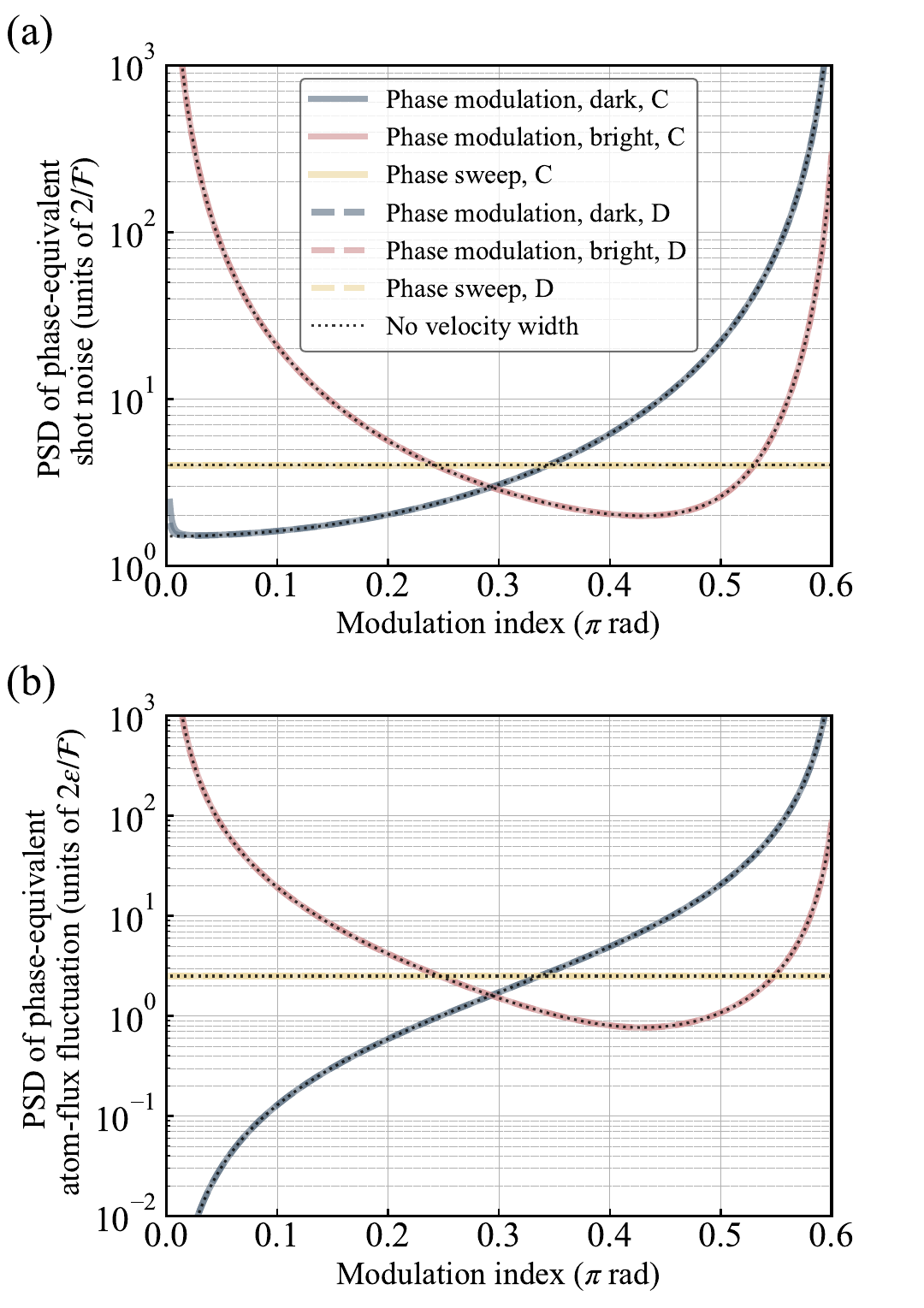}
        \caption{Power spectral density of phase-equivalent shot noise and atom-flux fluctuation in the case of a laser-cooled atomic beam.
        The results for ports C and D overlap with each other and the difference is hardly visible in the plots.
        The results for the case of fixed atomic velocity are also plotted with the black dotted lines for reference.
        They overlap in almost all the regions.}
        \label{fig:psd_cooled}
\end{figure}

We calculate the sensitivity for a laser-cooled atomic beam using realistic parameters to determine the effect of a finite velocity width.
We refer to recently reported experimental results~\cite{Kwolek:2021yqp} for a feasible velocity width with up-to-date cooling technology.
The longitudinal temperature of the $^{87}$Rb beam was \SI{15}{\micro K} and the mean velocity was \SI{10.75}{m/s}.
We assume a Gaussian velocity distribution.
The sensitivities calculated using these parameters are presented in Fig.~\ref{fig:psd_cooled}.
The sensitivity for the laser-cooled atomic beam overlaps with that of the zero-velocity width in almost all regions.
Therefore, we conclude that our calculation, which does not consider velocity width, is a good model for atomic beams laser cooled using recent cooling techniques.


\bibliography{ai_pm}

\begin{thebibliography}{42}%
\makeatletter
\providecommand \@ifxundefined [1]{%
 \@ifx{#1\undefined}
}%
\providecommand \@ifnum [1]{%
 \ifnum #1\expandafter \@firstoftwo
 \else \expandafter \@secondoftwo
 \fi
}%
\providecommand \@ifx [1]{%
 \ifx #1\expandafter \@firstoftwo
 \else \expandafter \@secondoftwo
 \fi
}%
\providecommand \natexlab [1]{#1}%
\providecommand \enquote  [1]{``#1''}%
\providecommand \bibnamefont  [1]{#1}%
\providecommand \bibfnamefont [1]{#1}%
\providecommand \citenamefont [1]{#1}%
\providecommand \href@noop [0]{\@secondoftwo}%
\providecommand \href [0]{\begingroup \@sanitize@url \@href}%
\providecommand \@href[1]{\@@startlink{#1}\@@href}%
\providecommand \@@href[1]{\endgroup#1\@@endlink}%
\providecommand \@sanitize@url [0]{\catcode `\\12\catcode `\$12\catcode
  `\&12\catcode `\#12\catcode `\^12\catcode `\_12\catcode `\%12\relax}%
\providecommand \@@startlink[1]{}%
\providecommand \@@endlink[0]{}%
\providecommand \url  [0]{\begingroup\@sanitize@url \@url }%
\providecommand \@url [1]{\endgroup\@href {#1}{\urlprefix }}%
\providecommand \urlprefix  [0]{URL }%
\providecommand \Eprint [0]{\href }%
\providecommand \doibase [0]{https://doi.org/}%
\providecommand \selectlanguage [0]{\@gobble}%
\providecommand \bibinfo  [0]{\@secondoftwo}%
\providecommand \bibfield  [0]{\@secondoftwo}%
\providecommand \translation [1]{[#1]}%
\providecommand \BibitemOpen [0]{}%
\providecommand \bibitemStop [0]{}%
\providecommand \bibitemNoStop [0]{.\EOS\space}%
\providecommand \EOS [0]{\spacefactor3000\relax}%
\providecommand \BibitemShut  [1]{\csname bibitem#1\endcsname}%
\let\auto@bib@innerbib\@empty
\bibitem [{\citenamefont {Kasevich}\ and\ \citenamefont
  {Chu}(1991)}]{Kasevich:1991zz}%
  \BibitemOpen
  \bibfield  {author} {\bibinfo {author} {\bibfnamefont {M.}~\bibnamefont
  {Kasevich}}\ and\ \bibinfo {author} {\bibfnamefont {S.}~\bibnamefont {Chu}},\
  }\bibfield  {title} {\bibinfo {title} {{Atomic interferometry using
  stimulated Raman transitions}},\ }\href
  {https://doi.org/10.1103/PhysRevLett.67.181} {\bibfield  {journal} {\bibinfo
  {journal} {Phys. Rev. Lett.}\ }\textbf {\bibinfo {volume} {67}},\ \bibinfo
  {pages} {181} (\bibinfo {year} {1991})}\BibitemShut {NoStop}%
\bibitem [{\citenamefont {Riehle}\ \emph {et~al.}(1991)\citenamefont {Riehle},
  \citenamefont {Kisters}, \citenamefont {Witte}, \citenamefont {Helmcke},\
  and\ \citenamefont {Bord\'e}}]{Riehle:1991}%
  \BibitemOpen
  \bibfield  {author} {\bibinfo {author} {\bibfnamefont {F.}~\bibnamefont
  {Riehle}}, \bibinfo {author} {\bibfnamefont {T.}~\bibnamefont {Kisters}},
  \bibinfo {author} {\bibfnamefont {A.}~\bibnamefont {Witte}}, \bibinfo
  {author} {\bibfnamefont {J.}~\bibnamefont {Helmcke}},\ and\ \bibinfo {author}
  {\bibfnamefont {C.~J.}\ \bibnamefont {Bord\'e}},\ }\bibfield  {title}
  {\bibinfo {title} {{Optical Ramsey spectroscopy in a rotating frame: Sagnac
  effect in a matter-wave interferometer}},\ }\href
  {https://doi.org/10.1103/PhysRevLett.67.177} {\bibfield  {journal} {\bibinfo
  {journal} {Phys. Rev. Lett.}\ }\textbf {\bibinfo {volume} {67}},\ \bibinfo
  {pages} {177} (\bibinfo {year} {1991})}\BibitemShut {NoStop}%
\bibitem [{\citenamefont {Kasevich}\ and\ \citenamefont
  {Chu}(1992)}]{Kasevich:1992yii}%
  \BibitemOpen
  \bibfield  {author} {\bibinfo {author} {\bibfnamefont {M.}~\bibnamefont
  {Kasevich}}\ and\ \bibinfo {author} {\bibfnamefont {S.}~\bibnamefont {Chu}},\
  }\bibfield  {title} {\bibinfo {title} {{Measurement of the gravitational
  acceleration of an atom with a light-pulse atom interferometer}},\ }\href
  {https://doi.org/10.1007/BF00325375} {\bibfield  {journal} {\bibinfo
  {journal} {Appl. Phys. B}\ }\textbf {\bibinfo {volume} {54}},\ \bibinfo
  {pages} {321} (\bibinfo {year} {1992})}\BibitemShut {NoStop}%
\bibitem [{\citenamefont {Geiger}\ \emph {et~al.}(2020)\citenamefont {Geiger},
  \citenamefont {Landragin}, \citenamefont {Merlet},\ and\ \citenamefont
  {Pereira Dos~Santos}}]{Geiger:2020aeq}%
  \BibitemOpen
  \bibfield  {author} {\bibinfo {author} {\bibfnamefont {R.}~\bibnamefont
  {Geiger}}, \bibinfo {author} {\bibfnamefont {A.}~\bibnamefont {Landragin}},
  \bibinfo {author} {\bibfnamefont {S.}~\bibnamefont {Merlet}},\ and\ \bibinfo
  {author} {\bibfnamefont {F.}~\bibnamefont {Pereira Dos~Santos}},\ }\bibfield
  {title} {\bibinfo {title} {{High-accuracy inertial measurements with
  cold-atom sensors}},\ }\href {https://doi.org/10.1116/5.0009093} {\bibfield
  {journal} {\bibinfo  {journal} {AVS Quantum Sci.}\ }\textbf {\bibinfo
  {volume} {2}},\ \bibinfo {pages} {024702} (\bibinfo {year}
  {2020})}\BibitemShut {NoStop}%
\bibitem [{\citenamefont {McGuinness}\ \emph {et~al.}(2012)\citenamefont
  {McGuinness}, \citenamefont {Rakholia},\ and\ \citenamefont
  {Biedermann}}]{McGuinness:2012}%
  \BibitemOpen
  \bibfield  {author} {\bibinfo {author} {\bibfnamefont {H.~J.}\ \bibnamefont
  {McGuinness}}, \bibinfo {author} {\bibfnamefont {A.~V.}\ \bibnamefont
  {Rakholia}},\ and\ \bibinfo {author} {\bibfnamefont {G.~W.}\ \bibnamefont
  {Biedermann}},\ }\bibfield  {title} {\bibinfo {title} {{High data-rate atom
  interferometer for measuring acceleration}},\ }\href
  {https://doi.org/10.1063/1.3673845} {\bibfield  {journal} {\bibinfo
  {journal} {Appl. Phys. Lett.}\ }\textbf {\bibinfo {volume} {100}},\ \bibinfo
  {pages} {011106} (\bibinfo {year} {2012})}\BibitemShut {NoStop}%
\bibitem [{\citenamefont {Lautier}\ \emph {et~al.}(2014)\citenamefont
  {Lautier}, \citenamefont {Volodimer}, \citenamefont {Hardin}, \citenamefont
  {Merlet}, \citenamefont {Lours}, \citenamefont {Pereira Dos~Santos},\ and\
  \citenamefont {Landragin}}]{Lautier:2014gaa}%
  \BibitemOpen
  \bibfield  {author} {\bibinfo {author} {\bibfnamefont {J.}~\bibnamefont
  {Lautier}}, \bibinfo {author} {\bibfnamefont {L.}~\bibnamefont {Volodimer}},
  \bibinfo {author} {\bibfnamefont {T.}~\bibnamefont {Hardin}}, \bibinfo
  {author} {\bibfnamefont {S.}~\bibnamefont {Merlet}}, \bibinfo {author}
  {\bibfnamefont {M.}~\bibnamefont {Lours}}, \bibinfo {author} {\bibfnamefont
  {F.}~\bibnamefont {Pereira Dos~Santos}},\ and\ \bibinfo {author}
  {\bibfnamefont {A.}~\bibnamefont {Landragin}},\ }\bibfield  {title} {\bibinfo
  {title} {{Hybridizing matter-wave and classical accelerometers}},\ }\href
  {https://doi.org/10.1063/1.4897358} {\bibfield  {journal} {\bibinfo
  {journal} {Appl. Phys. Lett.}\ }\textbf {\bibinfo {volume} {105}},\ \bibinfo
  {pages} {144102} (\bibinfo {year} {2014})}\BibitemShut {NoStop}%
\bibitem [{\citenamefont {Cheiney}\ \emph {et~al.}(2018)\citenamefont
  {Cheiney}, \citenamefont {Fouch\'e}, \citenamefont {Templier}, \citenamefont
  {Napolitano}, \citenamefont {Battelier}, \citenamefont {Bouyer},\ and\
  \citenamefont {Barrett}}]{Cheiney:2018}%
  \BibitemOpen
  \bibfield  {author} {\bibinfo {author} {\bibfnamefont {P.}~\bibnamefont
  {Cheiney}}, \bibinfo {author} {\bibfnamefont {L.}~\bibnamefont {Fouch\'e}},
  \bibinfo {author} {\bibfnamefont {S.}~\bibnamefont {Templier}}, \bibinfo
  {author} {\bibfnamefont {F.}~\bibnamefont {Napolitano}}, \bibinfo {author}
  {\bibfnamefont {B.}~\bibnamefont {Battelier}}, \bibinfo {author}
  {\bibfnamefont {P.}~\bibnamefont {Bouyer}},\ and\ \bibinfo {author}
  {\bibfnamefont {B.}~\bibnamefont {Barrett}},\ }\bibfield  {title} {\bibinfo
  {title} {{Navigation-compatible hybrid quantum accelerometer using a Kalman
  filter}},\ }\href {https://doi.org/10.1103/PhysRevApplied.10.034030}
  {\bibfield  {journal} {\bibinfo  {journal} {Phys. Rev. Appl.}\ }\textbf
  {\bibinfo {volume} {10}},\ \bibinfo {pages} {034030} (\bibinfo {year}
  {2018})}\BibitemShut {NoStop}%
\bibitem [{\citenamefont {Gustavson}\ \emph {et~al.}(2000)\citenamefont
  {Gustavson}, \citenamefont {Landragin},\ and\ \citenamefont
  {Kasevich}}]{Gustavson:2000}%
  \BibitemOpen
  \bibfield  {author} {\bibinfo {author} {\bibfnamefont {T.~L.}\ \bibnamefont
  {Gustavson}}, \bibinfo {author} {\bibfnamefont {A.}~\bibnamefont
  {Landragin}},\ and\ \bibinfo {author} {\bibfnamefont {M.~A.}\ \bibnamefont
  {Kasevich}},\ }\bibfield  {title} {\bibinfo {title} {{Rotation sensing with a
  dual atom-interferometer Sagnac gyroscope}},\ }\href
  {https://doi.org/10.1088/0264-9381/17/12/311} {\bibfield  {journal} {\bibinfo
   {journal} {Class. Quant. Grav.}\ }\textbf {\bibinfo {volume} {17}},\
  \bibinfo {pages} {2385} (\bibinfo {year} {2000})}\BibitemShut {NoStop}%
\bibitem [{\citenamefont {Durfee}\ \emph {et~al.}(2006)\citenamefont {Durfee},
  \citenamefont {Shaham},\ and\ \citenamefont {Kasevich}}]{Durfee:2005fna}%
  \BibitemOpen
  \bibfield  {author} {\bibinfo {author} {\bibfnamefont {D.~S.}\ \bibnamefont
  {Durfee}}, \bibinfo {author} {\bibfnamefont {Y.~K.}\ \bibnamefont {Shaham}},\
  and\ \bibinfo {author} {\bibfnamefont {M.~A.}\ \bibnamefont {Kasevich}},\
  }\bibfield  {title} {\bibinfo {title} {{Long-term stability of an
  area-reversible atom-interferometer Sagnac gyroscope}},\ }\href
  {https://doi.org/10.1103/PhysRevLett.97.240801} {\bibfield  {journal}
  {\bibinfo  {journal} {Phys. Rev. Lett.}\ }\textbf {\bibinfo {volume} {97}},\
  \bibinfo {pages} {240801} (\bibinfo {year} {2006})}\BibitemShut {NoStop}%
\bibitem [{\citenamefont {Savoie}\ \emph {et~al.}(2018)\citenamefont {Savoie},
  \citenamefont {Altorio}, \citenamefont {Fang}, \citenamefont {Sidorenkov},
  \citenamefont {Geiger},\ and\ \citenamefont {Landragin}}]{Savoie:2018guq}%
  \BibitemOpen
  \bibfield  {author} {\bibinfo {author} {\bibfnamefont {D.}~\bibnamefont
  {Savoie}}, \bibinfo {author} {\bibfnamefont {M.}~\bibnamefont {Altorio}},
  \bibinfo {author} {\bibfnamefont {B.}~\bibnamefont {Fang}}, \bibinfo {author}
  {\bibfnamefont {L.~A.}\ \bibnamefont {Sidorenkov}}, \bibinfo {author}
  {\bibfnamefont {R.}~\bibnamefont {Geiger}},\ and\ \bibinfo {author}
  {\bibfnamefont {A.}~\bibnamefont {Landragin}},\ }\bibfield  {title} {\bibinfo
  {title} {{Interleaved atom interferometry for high sensitivity inertial
  measurements}},\ }\href {https://doi.org/10.1126/sciadv.aau7948} {\bibfield
  {journal} {\bibinfo  {journal} {Sci. Adv.}\ }\textbf {\bibinfo {volume}
  {4}},\ \bibinfo {pages} {eaau7948} (\bibinfo {year} {2018})}\BibitemShut
  {NoStop}%
\bibitem [{\citenamefont {Avinadav}\ \emph {et~al.}(2020)\citenamefont
  {Avinadav}, \citenamefont {Yankelev}, \citenamefont {Shuker}, \citenamefont
  {Firstenberg},\ and\ \citenamefont {Davidson}}]{Avinadav:2020}%
  \BibitemOpen
  \bibfield  {author} {\bibinfo {author} {\bibfnamefont {C.}~\bibnamefont
  {Avinadav}}, \bibinfo {author} {\bibfnamefont {D.}~\bibnamefont {Yankelev}},
  \bibinfo {author} {\bibfnamefont {M.}~\bibnamefont {Shuker}}, \bibinfo
  {author} {\bibfnamefont {O.}~\bibnamefont {Firstenberg}},\ and\ \bibinfo
  {author} {\bibfnamefont {N.}~\bibnamefont {Davidson}},\ }\bibfield  {title}
  {\bibinfo {title} {{Rotation sensing with improved stability using
  point-source atom interferometry}},\ }\href
  {https://doi.org/10.1103/PhysRevA.102.013326} {\bibfield  {journal} {\bibinfo
   {journal} {Phys. Rev. A}\ }\textbf {\bibinfo {volume} {102}},\ \bibinfo
  {pages} {013326} (\bibinfo {year} {2020})}\BibitemShut {NoStop}%
\bibitem [{\citenamefont {Peters}\ \emph {et~al.}(2001)\citenamefont {Peters},
  \citenamefont {Chung},\ and\ \citenamefont {Chu}}]{Peters:2001}%
  \BibitemOpen
  \bibfield  {author} {\bibinfo {author} {\bibfnamefont {A.}~\bibnamefont
  {Peters}}, \bibinfo {author} {\bibfnamefont {K.~Y.}\ \bibnamefont {Chung}},\
  and\ \bibinfo {author} {\bibfnamefont {S.}~\bibnamefont {Chu}},\ }\bibfield
  {title} {\bibinfo {title} {{High-precision gravity measurements using atom
  interferometry}},\ }\href {https://doi.org/10.1088/0026-1394/38/1/4}
  {\bibfield  {journal} {\bibinfo  {journal} {Metrologia}\ }\textbf {\bibinfo
  {volume} {38}},\ \bibinfo {pages} {25} (\bibinfo {year} {2001})}\BibitemShut
  {NoStop}%
\bibitem [{\citenamefont {Bidel}\ \emph {et~al.}(2013)\citenamefont {Bidel},
  \citenamefont {Carraz}, \citenamefont {Charri\`ere}, \citenamefont {Cadoret},
  \citenamefont {Zahzam},\ and\ \citenamefont {Bresson}}]{Bidel:2013oja}%
  \BibitemOpen
  \bibfield  {author} {\bibinfo {author} {\bibfnamefont {Y.}~\bibnamefont
  {Bidel}}, \bibinfo {author} {\bibfnamefont {O.}~\bibnamefont {Carraz}},
  \bibinfo {author} {\bibfnamefont {R.}~\bibnamefont {Charri\`ere}}, \bibinfo
  {author} {\bibfnamefont {M.}~\bibnamefont {Cadoret}}, \bibinfo {author}
  {\bibfnamefont {N.}~\bibnamefont {Zahzam}},\ and\ \bibinfo {author}
  {\bibfnamefont {A.}~\bibnamefont {Bresson}},\ }\bibfield  {title} {\bibinfo
  {title} {{Compact cold atom gravimeter for field applications}},\ }\href
  {https://doi.org/10.1063/1.4801756} {\bibfield  {journal} {\bibinfo
  {journal} {Appl. Phys. Lett.}\ }\textbf {\bibinfo {volume} {102}},\ \bibinfo
  {pages} {144107} (\bibinfo {year} {2013})}\BibitemShut {NoStop}%
\bibitem [{\citenamefont {Wu}\ \emph {et~al.}(2019)\citenamefont {Wu},
  \citenamefont {Pagel}, \citenamefont {Malek}, \citenamefont {Nguyen},
  \citenamefont {Zi}, \citenamefont {Scheirer},\ and\ \citenamefont
  {M\"uller}}]{Wu:2019ikc}%
  \BibitemOpen
  \bibfield  {author} {\bibinfo {author} {\bibfnamefont {X.}~\bibnamefont
  {Wu}}, \bibinfo {author} {\bibfnamefont {Z.}~\bibnamefont {Pagel}}, \bibinfo
  {author} {\bibfnamefont {B.~S.}\ \bibnamefont {Malek}}, \bibinfo {author}
  {\bibfnamefont {T.~H.}\ \bibnamefont {Nguyen}}, \bibinfo {author}
  {\bibfnamefont {F.}~\bibnamefont {Zi}}, \bibinfo {author} {\bibfnamefont
  {D.~S.}\ \bibnamefont {Scheirer}},\ and\ \bibinfo {author} {\bibfnamefont
  {H.}~\bibnamefont {M\"uller}},\ }\bibfield  {title} {\bibinfo {title}
  {{Gravity surveys using a mobile atom interferometer}},\ }\href
  {https://doi.org/10.1126/sciadv.aax0800} {\bibfield  {journal} {\bibinfo
  {journal} {Sci. Adv.}\ }\textbf {\bibinfo {volume} {5}},\ \bibinfo {pages}
  {eaax0800} (\bibinfo {year} {2019})}\BibitemShut {NoStop}%
\bibitem [{\citenamefont {Snadden}\ \emph {et~al.}(1998)\citenamefont
  {Snadden}, \citenamefont {McGuirk}, \citenamefont {Bouyer}, \citenamefont
  {Haritos},\ and\ \citenamefont {Kasevich}}]{Snadden:1998zz}%
  \BibitemOpen
  \bibfield  {author} {\bibinfo {author} {\bibfnamefont {M.~J.}\ \bibnamefont
  {Snadden}}, \bibinfo {author} {\bibfnamefont {J.~M.}\ \bibnamefont
  {McGuirk}}, \bibinfo {author} {\bibfnamefont {P.}~\bibnamefont {Bouyer}},
  \bibinfo {author} {\bibfnamefont {K.~G.}\ \bibnamefont {Haritos}},\ and\
  \bibinfo {author} {\bibfnamefont {M.~A.}\ \bibnamefont {Kasevich}},\
  }\bibfield  {title} {\bibinfo {title} {{Measurement of the earth's gravity
  gradient with an atom interferometer-based gravity gradiometer}},\ }\href
  {https://doi.org/10.1103/PhysRevLett.81.971} {\bibfield  {journal} {\bibinfo
  {journal} {Phys. Rev. Lett.}\ }\textbf {\bibinfo {volume} {81}},\ \bibinfo
  {pages} {971} (\bibinfo {year} {1998})}\BibitemShut {NoStop}%
\bibitem [{\citenamefont {Biedermann}\ \emph {et~al.}(2015)\citenamefont
  {Biedermann}, \citenamefont {Wu}, \citenamefont {Deslauriers}, \citenamefont
  {Roy}, \citenamefont {Mahadeswaraswamy},\ and\ \citenamefont
  {Kasevich}}]{Biedermann:2014jya}%
  \BibitemOpen
  \bibfield  {author} {\bibinfo {author} {\bibfnamefont {G.~W.}\ \bibnamefont
  {Biedermann}}, \bibinfo {author} {\bibfnamefont {X.}~\bibnamefont {Wu}},
  \bibinfo {author} {\bibfnamefont {L.}~\bibnamefont {Deslauriers}}, \bibinfo
  {author} {\bibfnamefont {S.}~\bibnamefont {Roy}}, \bibinfo {author}
  {\bibfnamefont {C.}~\bibnamefont {Mahadeswaraswamy}},\ and\ \bibinfo {author}
  {\bibfnamefont {M.~A.}\ \bibnamefont {Kasevich}},\ }\bibfield  {title}
  {\bibinfo {title} {{Testing gravity with cold-atom interferometers}},\ }\href
  {https://doi.org/10.1103/PhysRevA.91.033629} {\bibfield  {journal} {\bibinfo
  {journal} {Phys. Rev. A}\ }\textbf {\bibinfo {volume} {91}},\ \bibinfo
  {pages} {033629} (\bibinfo {year} {2015})}\BibitemShut {NoStop}%
\bibitem [{\citenamefont {Fray}\ \emph {et~al.}(2004)\citenamefont {Fray},
  \citenamefont {Diez}, \citenamefont {H{\"a}nsch},\ and\ \citenamefont
  {Weitz}}]{Fray:2004zs}%
  \BibitemOpen
  \bibfield  {author} {\bibinfo {author} {\bibfnamefont {S.}~\bibnamefont
  {Fray}}, \bibinfo {author} {\bibfnamefont {C.~A.}\ \bibnamefont {Diez}},
  \bibinfo {author} {\bibfnamefont {T.~W.}\ \bibnamefont {H{\"a}nsch}},\ and\
  \bibinfo {author} {\bibfnamefont {M.}~\bibnamefont {Weitz}},\ }\bibfield
  {title} {\bibinfo {title} {{Atomic interferometer with amplitude gratings of
  light and its applications to atom based tests of the equivalence
  principle}},\ }\href {https://doi.org/10.1103/PhysRevLett.93.240404}
  {\bibfield  {journal} {\bibinfo  {journal} {Phys. Rev. Lett.}\ }\textbf
  {\bibinfo {volume} {93}},\ \bibinfo {pages} {240404} (\bibinfo {year}
  {2004})}\BibitemShut {NoStop}%
\bibitem [{\citenamefont {Tarallo}\ \emph {et~al.}(2014)\citenamefont
  {Tarallo}, \citenamefont {Mazzoni}, \citenamefont {Poli}, \citenamefont
  {Sutyrin}, \citenamefont {Zhang},\ and\ \citenamefont
  {Tino}}]{Tarallo:2014oaa}%
  \BibitemOpen
  \bibfield  {author} {\bibinfo {author} {\bibfnamefont {M.~G.}\ \bibnamefont
  {Tarallo}}, \bibinfo {author} {\bibfnamefont {T.}~\bibnamefont {Mazzoni}},
  \bibinfo {author} {\bibfnamefont {N.}~\bibnamefont {Poli}}, \bibinfo {author}
  {\bibfnamefont {D.~V.}\ \bibnamefont {Sutyrin}}, \bibinfo {author}
  {\bibfnamefont {X.}~\bibnamefont {Zhang}},\ and\ \bibinfo {author}
  {\bibfnamefont {G.~M.}\ \bibnamefont {Tino}},\ }\bibfield  {title} {\bibinfo
  {title} {{Test of Einstein Equivalence Principle for 0-spin and
  half-integer-spin atoms: Search for spin-gravity coupling effects}},\ }\href
  {https://doi.org/10.1103/PhysRevLett.113.023005} {\bibfield  {journal}
  {\bibinfo  {journal} {Phys. Rev. Lett.}\ }\textbf {\bibinfo {volume} {113}},\
  \bibinfo {pages} {023005} (\bibinfo {year} {2014})}\BibitemShut {NoStop}%
\bibitem [{\citenamefont {Zhou}\ \emph {et~al.}(2015)\citenamefont {Zhou} \emph
  {et~al.}}]{Zhou:2015pna}%
  \BibitemOpen
  \bibfield  {author} {\bibinfo {author} {\bibfnamefont {L.}~\bibnamefont
  {Zhou}} \emph {et~al.},\ }\bibfield  {title} {\bibinfo {title} {{Test of
  Equivalence Principle at $10^{-8}$ Level by a Dual-species Double-diffraction
  Raman Atom Interferometer}},\ }\href
  {https://doi.org/10.1103/PhysRevLett.115.013004} {\bibfield  {journal}
  {\bibinfo  {journal} {Phys. Rev. Lett.}\ }\textbf {\bibinfo {volume} {115}},\
  \bibinfo {pages} {013004} (\bibinfo {year} {2015})}\BibitemShut {NoStop}%
\bibitem [{\citenamefont {Rosi}\ \emph {et~al.}(2017)\citenamefont {Rosi},
  \citenamefont {D'Amico}, \citenamefont {Cacciapuoti}, \citenamefont
  {Sorrentino}, \citenamefont {Prevedelli}, \citenamefont {Zych}, \citenamefont
  {Brukner},\ and\ \citenamefont {Tino}}]{Rosi:2017ieh}%
  \BibitemOpen
  \bibfield  {author} {\bibinfo {author} {\bibfnamefont {G.}~\bibnamefont
  {Rosi}}, \bibinfo {author} {\bibfnamefont {G.}~\bibnamefont {D'Amico}},
  \bibinfo {author} {\bibfnamefont {L.}~\bibnamefont {Cacciapuoti}}, \bibinfo
  {author} {\bibfnamefont {F.}~\bibnamefont {Sorrentino}}, \bibinfo {author}
  {\bibfnamefont {M.}~\bibnamefont {Prevedelli}}, \bibinfo {author}
  {\bibfnamefont {M.}~\bibnamefont {Zych}}, \bibinfo {author} {\bibfnamefont
  {C.}~\bibnamefont {Brukner}},\ and\ \bibinfo {author} {\bibfnamefont {G.~M.}\
  \bibnamefont {Tino}},\ }\bibfield  {title} {\bibinfo {title} {{Quantum test
  of the equivalence principle for atoms in superpositions of internal energy
  eigenstates}},\ }\href {https://doi.org/10.1038/ncomms15529} {\bibfield
  {journal} {\bibinfo  {journal} {Nature Commun.}\ }\textbf {\bibinfo {volume}
  {8}},\ \bibinfo {pages} {5529} (\bibinfo {year} {2017})}\BibitemShut
  {NoStop}%
\bibitem [{\citenamefont {Tino}\ and\ \citenamefont
  {Vetrano}(2011)}]{Tino:2011zz}%
  \BibitemOpen
  \bibfield  {author} {\bibinfo {author} {\bibfnamefont {G.~M.}\ \bibnamefont
  {Tino}}\ and\ \bibinfo {author} {\bibfnamefont {F.}~\bibnamefont {Vetrano}},\
  }\bibfield  {title} {\bibinfo {title} {{Atom interferometers for
  gravitational wave detection: A look at a `simple' configuration}},\ }\href
  {https://doi.org/10.1007/s10714-010-1139-5} {\bibfield  {journal} {\bibinfo
  {journal} {Gen. Relat. Gravit.}\ }\textbf {\bibinfo {volume} {43}},\ \bibinfo
  {pages} {2037} (\bibinfo {year} {2011})}\BibitemShut {NoStop}%
\bibitem [{\citenamefont {Dimopoulos}\ \emph {et~al.}(2009)\citenamefont
  {Dimopoulos}, \citenamefont {Graham}, \citenamefont {Hogan}, \citenamefont
  {Kasevich},\ and\ \citenamefont {Rajendran}}]{Dimopoulos:2007cj}%
  \BibitemOpen
  \bibfield  {author} {\bibinfo {author} {\bibfnamefont {S.}~\bibnamefont
  {Dimopoulos}}, \bibinfo {author} {\bibfnamefont {P.~W.}\ \bibnamefont
  {Graham}}, \bibinfo {author} {\bibfnamefont {J.~M.}\ \bibnamefont {Hogan}},
  \bibinfo {author} {\bibfnamefont {M.~A.}\ \bibnamefont {Kasevich}},\ and\
  \bibinfo {author} {\bibfnamefont {S.}~\bibnamefont {Rajendran}},\ }\bibfield
  {title} {\bibinfo {title} {{Gravitational wave detection with atom
  interferometry}},\ }\href {https://doi.org/10.1016/j.physletb.2009.06.011}
  {\bibfield  {journal} {\bibinfo  {journal} {Phys. Lett. B}\ }\textbf
  {\bibinfo {volume} {678}},\ \bibinfo {pages} {37} (\bibinfo {year}
  {2009})}\BibitemShut {NoStop}%
\bibitem [{\citenamefont {Canuel}\ \emph {et~al.}(2018)\citenamefont {Canuel}
  \emph {et~al.}}]{Canuel:2017rrp}%
  \BibitemOpen
  \bibfield  {author} {\bibinfo {author} {\bibfnamefont {B.}~\bibnamefont
  {Canuel}} \emph {et~al.},\ }\bibfield  {title} {\bibinfo {title} {{Exploring
  gravity with the MIGA large scale atom interferometer}},\ }\href
  {https://doi.org/10.1038/s41598-018-32165-z} {\bibfield  {journal} {\bibinfo
  {journal} {Sci. Rep.}\ }\textbf {\bibinfo {volume} {8}},\ \bibinfo {pages}
  {14064} (\bibinfo {year} {2018})}\BibitemShut {NoStop}%
\bibitem [{\citenamefont {Abe}\ \emph {et~al.}(2021)\citenamefont {Abe} \emph
  {et~al.}}]{MAGIS-100:2021etm}%
  \BibitemOpen
  \bibfield  {author} {\bibinfo {author} {\bibfnamefont {M.}~\bibnamefont
  {Abe}} \emph {et~al.},\ }\bibfield  {title} {\bibinfo {title} {{Matter-wave
  atomic gradiometer interferometric sensor (MAGIS-100)}},\ }\href
  {https://doi.org/10.1088/2058-9565/abf719} {\bibfield  {journal} {\bibinfo
  {journal} {Quantum Sci. Technol.}\ }\textbf {\bibinfo {volume} {6}},\
  \bibinfo {pages} {044003} (\bibinfo {year} {2021})}\BibitemShut {NoStop}%
\bibitem [{\citenamefont {Joyet}\ \emph {et~al.}(2012)\citenamefont {Joyet},
  \citenamefont {Domenico},\ and\ \citenamefont {Thomann}}]{Joyet:2012}%
  \BibitemOpen
  \bibfield  {author} {\bibinfo {author} {\bibfnamefont {A.}~\bibnamefont
  {Joyet}}, \bibinfo {author} {\bibfnamefont {G.~D.}\ \bibnamefont
  {Domenico}},\ and\ \bibinfo {author} {\bibfnamefont {P.}~\bibnamefont
  {Thomann}},\ }\bibfield  {title} {\bibinfo {title} {{Theoretical analysis of
  aliasing noises in cold atom Mach-Zehnder interferometers}},\ }\href
  {https://doi.org/10.1140/epjd/e2012-20401-6} {\bibfield  {journal} {\bibinfo
  {journal} {Eur. Phys. J. D}\ }\textbf {\bibinfo {volume} {66}},\ \bibinfo
  {pages} {61} (\bibinfo {year} {2012})}\BibitemShut {NoStop}%
\bibitem [{\citenamefont {Lenef}\ \emph {et~al.}(1997)\citenamefont {Lenef},
  \citenamefont {Hammond}, \citenamefont {Smith}, \citenamefont {Chapman},
  \citenamefont {Rubenstein},\ and\ \citenamefont {Pritchard}}]{Lenef:1997}%
  \BibitemOpen
  \bibfield  {author} {\bibinfo {author} {\bibfnamefont {A.}~\bibnamefont
  {Lenef}}, \bibinfo {author} {\bibfnamefont {T.~D.}\ \bibnamefont {Hammond}},
  \bibinfo {author} {\bibfnamefont {E.~T.}\ \bibnamefont {Smith}}, \bibinfo
  {author} {\bibfnamefont {M.~S.}\ \bibnamefont {Chapman}}, \bibinfo {author}
  {\bibfnamefont {R.~A.}\ \bibnamefont {Rubenstein}},\ and\ \bibinfo {author}
  {\bibfnamefont {D.~E.}\ \bibnamefont {Pritchard}},\ }\bibfield  {title}
  {\bibinfo {title} {{Rotation sensing with an atom interferometer}},\ }\href
  {https://doi.org/10.1103/PhysRevLett.78.760} {\bibfield  {journal} {\bibinfo
  {journal} {Phys. Rev. Lett.}\ }\textbf {\bibinfo {volume} {78}},\ \bibinfo
  {pages} {760} (\bibinfo {year} {1997})}\BibitemShut {NoStop}%
\bibitem [{\citenamefont {Xue}\ \emph {et~al.}(2015)\citenamefont {Xue},
  \citenamefont {Feng}, \citenamefont {Chen}, \citenamefont {Wang},
  \citenamefont {Yan}, \citenamefont {Jiang},\ and\ \citenamefont
  {Zhou}}]{Xue:2015}%
  \BibitemOpen
  \bibfield  {author} {\bibinfo {author} {\bibfnamefont {H.}~\bibnamefont
  {Xue}}, \bibinfo {author} {\bibfnamefont {Y.}~\bibnamefont {Feng}}, \bibinfo
  {author} {\bibfnamefont {S.}~\bibnamefont {Chen}}, \bibinfo {author}
  {\bibfnamefont {X.}~\bibnamefont {Wang}}, \bibinfo {author} {\bibfnamefont
  {X.}~\bibnamefont {Yan}}, \bibinfo {author} {\bibfnamefont {Z.}~\bibnamefont
  {Jiang}},\ and\ \bibinfo {author} {\bibfnamefont {Z.}~\bibnamefont {Zhou}},\
  }\bibfield  {title} {\bibinfo {title} {{A continuous cold atomic beam
  interferometer}},\ }\href {https://doi.org/10.1063/1.4913711} {\bibfield
  {journal} {\bibinfo  {journal} {J. Appl. Phys.}\ }\textbf {\bibinfo {volume}
  {117}},\ \bibinfo {pages} {094901} (\bibinfo {year} {2015})}\BibitemShut
  {NoStop}%
\bibitem [{\citenamefont {Kwolek}\ \emph {et~al.}(2020)\citenamefont {Kwolek},
  \citenamefont {Fancher}, \citenamefont {Bashkansky},\ and\ \citenamefont
  {Black}}]{Kwolek:2020}%
  \BibitemOpen
  \bibfield  {author} {\bibinfo {author} {\bibfnamefont {J.~M.}\ \bibnamefont
  {Kwolek}}, \bibinfo {author} {\bibfnamefont {C.~T.}\ \bibnamefont {Fancher}},
  \bibinfo {author} {\bibfnamefont {M.}~\bibnamefont {Bashkansky}},\ and\
  \bibinfo {author} {\bibfnamefont {A.~T.}\ \bibnamefont {Black}},\ }\bibfield
  {title} {\bibinfo {title} {{Three-dimensional cooling of an atom-beam source
  for high-contrast atom interferometry}},\ }\href
  {https://doi.org/10.1103/PhysRevApplied.13.044057} {\bibfield  {journal}
  {\bibinfo  {journal} {Phys. Rev. Appl.}\ }\textbf {\bibinfo {volume} {13}},\
  \bibinfo {pages} {044057} (\bibinfo {year} {2020})}\BibitemShut {NoStop}%
\bibitem [{\citenamefont {Kwolek}\ and\ \citenamefont
  {Black}(2022)}]{Kwolek:2021yqp}%
  \BibitemOpen
  \bibfield  {author} {\bibinfo {author} {\bibfnamefont {J.~M.}\ \bibnamefont
  {Kwolek}}\ and\ \bibinfo {author} {\bibfnamefont {A.~T.}\ \bibnamefont
  {Black}},\ }\bibfield  {title} {\bibinfo {title} {{Continuous
  sub-Doppler-cooled atomic beam interferometer for inertial sensing}},\ }\href
  {https://doi.org/10.1103/PhysRevApplied.17.024061} {\bibfield  {journal}
  {\bibinfo  {journal} {Phys. Rev. Applied}\ }\textbf {\bibinfo {volume}
  {17}},\ \bibinfo {pages} {024061} (\bibinfo {year} {2022})}\BibitemShut
  {NoStop}%
\bibitem [{\citenamefont {Meng}\ \emph {et~al.}(2024)\citenamefont {Meng},
  \citenamefont {Yan}, \citenamefont {Wang}, \citenamefont {Li}, \citenamefont
  {Xue},\ and\ \citenamefont {Feng}}]{Meng:2022cbn}%
  \BibitemOpen
  \bibfield  {author} {\bibinfo {author} {\bibfnamefont {Z.-X.}\ \bibnamefont
  {Meng}}, \bibinfo {author} {\bibfnamefont {P.-Q.}\ \bibnamefont {Yan}},
  \bibinfo {author} {\bibfnamefont {S.-Z.}\ \bibnamefont {Wang}}, \bibinfo
  {author} {\bibfnamefont {X.-J.}\ \bibnamefont {Li}}, \bibinfo {author}
  {\bibfnamefont {H.-B.}\ \bibnamefont {Xue}},\ and\ \bibinfo {author}
  {\bibfnamefont {Y.-Y.}\ \bibnamefont {Feng}},\ }\bibfield  {title} {\bibinfo
  {title} {{Closed-loop dual-atom-interferometer inertial sensor with
  continuous cold atomic beams}},\ }\href
  {https://doi.org/10.1103/PhysRevApplied.21.034050} {\bibfield  {journal}
  {\bibinfo  {journal} {Phys. Rev. Appl.}\ }\textbf {\bibinfo {volume} {21}},\
  \bibinfo {pages} {034050} (\bibinfo {year} {2024})}\BibitemShut {NoStop}%
\bibitem [{\citenamefont {Rakholia}\ \emph {et~al.}(2014)\citenamefont
  {Rakholia}, \citenamefont {McGuinness},\ and\ \citenamefont
  {Biedermann}}]{Rakholia:2014}%
  \BibitemOpen
  \bibfield  {author} {\bibinfo {author} {\bibfnamefont {A.~V.}\ \bibnamefont
  {Rakholia}}, \bibinfo {author} {\bibfnamefont {H.~J.}\ \bibnamefont
  {McGuinness}},\ and\ \bibinfo {author} {\bibfnamefont {G.~W.}\ \bibnamefont
  {Biedermann}},\ }\bibfield  {title} {\bibinfo {title} {{Dual-axis
  high-data-rate atom interferometer via cold ensemble exchange}},\ }\href
  {https://doi.org/10.1103/PhysRevApplied.2.054012} {\bibfield  {journal}
  {\bibinfo  {journal} {Phys. Rev. Appl.}\ }\textbf {\bibinfo {volume} {2}},\
  \bibinfo {pages} {054012} (\bibinfo {year} {2014})}\BibitemShut {NoStop}%
\bibitem [{\citenamefont {Sato}\ \emph {et~al.}(2024)\citenamefont {Sato},
  \citenamefont {Nishimura}, \citenamefont {Kaku}, \citenamefont {Otabe},
  \citenamefont {Kawasaki}, \citenamefont {Hosoya},\ and\ \citenamefont
  {Kozuma}}]{Sato:2023}%
  \BibitemOpen
  \bibfield  {author} {\bibinfo {author} {\bibfnamefont {T.}~\bibnamefont
  {Sato}}, \bibinfo {author} {\bibfnamefont {N.}~\bibnamefont {Nishimura}},
  \bibinfo {author} {\bibfnamefont {N.}~\bibnamefont {Kaku}}, \bibinfo {author}
  {\bibfnamefont {S.}~\bibnamefont {Otabe}}, \bibinfo {author} {\bibfnamefont
  {T.}~\bibnamefont {Kawasaki}}, \bibinfo {author} {\bibfnamefont
  {T.}~\bibnamefont {Hosoya}},\ and\ \bibinfo {author} {\bibfnamefont
  {M.}~\bibnamefont {Kozuma}},\ }\bibfield  {title} {\bibinfo {title}
  {{Closed-loop measurement in atom interferometer gyroscope with
  velocity-dependent phase dispersion compensation}},\ }\Eprint
  {https://arxiv.org/abs/2407.05696} {arXiv:2407.05696 [physics.atom-ph]}
  (\bibinfo {year} {2024})\BibitemShut {NoStop}%
\bibitem [{\citenamefont {Itano}\ \emph {et~al.}(1993)\citenamefont {Itano},
  \citenamefont {Bergquist}, \citenamefont {Bollinger}, \citenamefont
  {Gilligan}, \citenamefont {Heinzen}, \citenamefont {Moore}, \citenamefont
  {Raizen},\ and\ \citenamefont {Wineland}}]{Itano:1993}%
  \BibitemOpen
  \bibfield  {author} {\bibinfo {author} {\bibfnamefont {W.~M.}\ \bibnamefont
  {Itano}}, \bibinfo {author} {\bibfnamefont {J.~C.}\ \bibnamefont
  {Bergquist}}, \bibinfo {author} {\bibfnamefont {J.~J.}\ \bibnamefont
  {Bollinger}}, \bibinfo {author} {\bibfnamefont {J.~M.}\ \bibnamefont
  {Gilligan}}, \bibinfo {author} {\bibfnamefont {D.~J.}\ \bibnamefont
  {Heinzen}}, \bibinfo {author} {\bibfnamefont {F.~L.}\ \bibnamefont {Moore}},
  \bibinfo {author} {\bibfnamefont {M.~G.}\ \bibnamefont {Raizen}},\ and\
  \bibinfo {author} {\bibfnamefont {D.~J.}\ \bibnamefont {Wineland}},\
  }\bibfield  {title} {\bibinfo {title} {{Quantum projection noise: Population
  fluctuations in two-level systems}},\ }\href
  {https://doi.org/10.1103/PhysRevA.47.3554} {\bibfield  {journal} {\bibinfo
  {journal} {Phys. Rev. A}\ }\textbf {\bibinfo {volume} {47}},\ \bibinfo
  {pages} {3554} (\bibinfo {year} {1993})}\BibitemShut {NoStop}%
\bibitem [{\citenamefont {Orzel}\ \emph {et~al.}(2001)\citenamefont {Orzel},
  \citenamefont {Tuchman}, \citenamefont {Fenselau}, \citenamefont {Yasuda},\
  and\ \citenamefont {Kasevich}}]{Orzel2001}%
  \BibitemOpen
  \bibfield  {author} {\bibinfo {author} {\bibfnamefont {C.}~\bibnamefont
  {Orzel}}, \bibinfo {author} {\bibfnamefont {A.~K.}\ \bibnamefont {Tuchman}},
  \bibinfo {author} {\bibfnamefont {M.~L.}\ \bibnamefont {Fenselau}}, \bibinfo
  {author} {\bibfnamefont {M.}~\bibnamefont {Yasuda}},\ and\ \bibinfo {author}
  {\bibfnamefont {M.~A.}\ \bibnamefont {Kasevich}},\ }\bibfield  {title}
  {\bibinfo {title} {{Squeezed states in a Bose-Einstein condensate}},\ }\href
  {https://doi.org/10.1126/science.1058149} {\bibfield  {journal} {\bibinfo
  {journal} {Science}\ }\textbf {\bibinfo {volume} {291}},\ \bibinfo {pages}
  {2386} (\bibinfo {year} {2001})}\BibitemShut {NoStop}%
\bibitem [{\citenamefont {Caves}\ and\ \citenamefont
  {Schumaker}(1985)}]{Caves:1985zz}%
  \BibitemOpen
  \bibfield  {author} {\bibinfo {author} {\bibfnamefont {C.~M.}\ \bibnamefont
  {Caves}}\ and\ \bibinfo {author} {\bibfnamefont {B.~L.}\ \bibnamefont
  {Schumaker}},\ }\bibfield  {title} {\bibinfo {title} {{New formalism for
  two-photon quantum optics. I. Quadrature phases and squeezed states}},\
  }\href {https://doi.org/10.1103/PhysRevA.31.3068} {\bibfield  {journal}
  {\bibinfo  {journal} {Phys. Rev. A}\ }\textbf {\bibinfo {volume} {31}},\
  \bibinfo {pages} {3068} (\bibinfo {year} {1985})}\BibitemShut {NoStop}%
\bibitem [{\citenamefont {Schumaker}\ and\ \citenamefont
  {Caves}(1985)}]{Schumaker:1985zz}%
  \BibitemOpen
  \bibfield  {author} {\bibinfo {author} {\bibfnamefont {B.~L.}\ \bibnamefont
  {Schumaker}}\ and\ \bibinfo {author} {\bibfnamefont {C.~M.}\ \bibnamefont
  {Caves}},\ }\bibfield  {title} {\bibinfo {title} {{New formalism for
  two-photon quantum optics. II. Mathematical foundation and compact
  notation}},\ }\href {https://doi.org/10.1103/PhysRevA.31.3093} {\bibfield
  {journal} {\bibinfo  {journal} {Phys. Rev. A}\ }\textbf {\bibinfo {volume}
  {31}},\ \bibinfo {pages} {3093} (\bibinfo {year} {1985})}\BibitemShut
  {NoStop}%
\bibitem [{\citenamefont {Corbitt}\ \emph {et~al.}(2005)\citenamefont
  {Corbitt}, \citenamefont {Chen},\ and\ \citenamefont
  {Mavalvala}}]{Corbitt:2005qv}%
  \BibitemOpen
  \bibfield  {author} {\bibinfo {author} {\bibfnamefont {T.}~\bibnamefont
  {Corbitt}}, \bibinfo {author} {\bibfnamefont {Y.}~\bibnamefont {Chen}},\ and\
  \bibinfo {author} {\bibfnamefont {N.}~\bibnamefont {Mavalvala}},\ }\bibfield
  {title} {\bibinfo {title} {{Mathematical framework for simulation of quantum
  fields in complex interferometers using the two-photon formalism}},\ }\href
  {https://doi.org/10.1103/PhysRevA.72.013818} {\bibfield  {journal} {\bibinfo
  {journal} {Phys. Rev. A}\ }\textbf {\bibinfo {volume} {72}},\ \bibinfo
  {pages} {013818} (\bibinfo {year} {2005})}\BibitemShut {NoStop}%
\bibitem [{\citenamefont {Buonanno}\ \emph {et~al.}(2003)\citenamefont
  {Buonanno}, \citenamefont {Chen},\ and\ \citenamefont
  {Mavalvala}}]{Buonanno:2003ch}%
  \BibitemOpen
  \bibfield  {author} {\bibinfo {author} {\bibfnamefont {A.}~\bibnamefont
  {Buonanno}}, \bibinfo {author} {\bibfnamefont {Y.}~\bibnamefont {Chen}},\
  and\ \bibinfo {author} {\bibfnamefont {N.}~\bibnamefont {Mavalvala}},\
  }\bibfield  {title} {\bibinfo {title} {{Quantum noise in laser interferometer
  gravitational wave detectors with a heterodyne readout scheme}},\ }\href
  {https://doi.org/10.1103/PhysRevD.67.122005} {\bibfield  {journal} {\bibinfo
  {journal} {Phys. Rev. D}\ }\textbf {\bibinfo {volume} {67}},\ \bibinfo
  {pages} {122005} (\bibinfo {year} {2003})}\BibitemShut {NoStop}%
\bibitem [{\citenamefont {Ramsey}(1986)}]{Ramsey:1986}%
  \BibitemOpen
  \bibfield  {author} {\bibinfo {author} {\bibfnamefont {N.~F.}\ \bibnamefont
  {Ramsey}},\ }\href@noop {} {\emph {\bibinfo {title} {{Molecular Beams}}}}\
  (\bibinfo  {publisher} {Oxford University Press},\ \bibinfo {address}
  {Oxford},\ \bibinfo {year} {1986})\ \bibinfo {note} {{\unskip{},
  Sec.~II.3.2}}\BibitemShut {NoStop}%
\bibitem [{\citenamefont {B\"ohm}\ \emph {et~al.}(1983)\citenamefont {B\"ohm},
  \citenamefont {Marten}, \citenamefont {Weidel},\ and\ \citenamefont
  {Petermann}}]{Bohm:1983}%
  \BibitemOpen
  \bibfield  {author} {\bibinfo {author} {\bibfnamefont {K.}~\bibnamefont
  {B\"ohm}}, \bibinfo {author} {\bibfnamefont {P.}~\bibnamefont {Marten}},
  \bibinfo {author} {\bibfnamefont {E.}~\bibnamefont {Weidel}},\ and\ \bibinfo
  {author} {\bibfnamefont {K.}~\bibnamefont {Petermann}},\ }\bibfield  {title}
  {\bibinfo {title} {{Direct rotation-rate detection with a fibre-optic gyro by
  using digital data processing}},\ }\href
  {https://doi.org/10.1049/el:19830677} {\bibfield  {journal} {\bibinfo
  {journal} {Electron. Lett.}\ }\textbf {\bibinfo {volume} {19}},\ \bibinfo
  {pages} {997} (\bibinfo {year} {1983})}\BibitemShut {NoStop}%
\bibitem [{\citenamefont {Miranda}\ \emph {et~al.}(2023)\citenamefont
  {Miranda}, \citenamefont {Takei}, \citenamefont {Miyazawa},\ and\
  \citenamefont {Kozuma}}]{Miranda:2023}%
  \BibitemOpen
  \bibfield  {author} {\bibinfo {author} {\bibfnamefont {M.}~\bibnamefont
  {Miranda}}, \bibinfo {author} {\bibfnamefont {N.}~\bibnamefont {Takei}},
  \bibinfo {author} {\bibfnamefont {Y.}~\bibnamefont {Miyazawa}},\ and\
  \bibinfo {author} {\bibfnamefont {M.}~\bibnamefont {Kozuma}},\ }\bibfield
  {title} {\bibinfo {title} {{Multi-harmonic modulation in a fiber-optic
  gyroscope}},\ }\href {https://doi.org/10.3390/s23094442} {\bibfield
  {journal} {\bibinfo  {journal} {Sensors}\ }\textbf {\bibinfo {volume} {23}},\
  \bibinfo {pages} {4442} (\bibinfo {year} {2023})}\BibitemShut {NoStop}%
\bibitem [{\citenamefont {Somiya}(2003)}]{Somiya:2002tx}%
  \BibitemOpen
  \bibfield  {author} {\bibinfo {author} {\bibfnamefont {K.}~\bibnamefont
  {Somiya}},\ }\bibfield  {title} {\bibinfo {title} {{New photodetection method
  using unbalanced sidebands for squeezed quantum noise in gravitational wave
  interferometer}},\ }\href {https://doi.org/10.1103/PhysRevD.67.122001}
  {\bibfield  {journal} {\bibinfo  {journal} {Phys. Rev. D}\ }\textbf {\bibinfo
  {volume} {67}},\ \bibinfo {pages} {122001} (\bibinfo {year}
  {2003})}\BibitemShut {NoStop}%
\end{thebibliography}%

\end{document}